\documentclass[prb,showpacs,twocolumn,superscriptaddress,floatfix]{revtex4}
\usepackage{hyperref,fontenc,times,amsmath,amsfonts,amssymb,graphicx,epsfig,color,bbm,bm,units,mathtools,hhline}
\usepackage{float}
\usepackage{placeins}
\usepackage{tikz}
\usepackage{threeparttable}
\usepackage{natbib}
\usepackage{relsize}
\usepackage{graphicx}
\usepackage{calc}

\newlength{\depthofsumsign}
\DeclarePairedDelimiterX\braket[2]{\langle}{\rangle}{#1\delimsize\vert}
\DeclarePairedDelimiterX\braxket[3]{\langle}{\rangle}{%
	#1\,\delimsize\vert\,#2\,\delimsize\vert\,#3}

\makeatletter
\renewcommand*\env@cases[1][1.2]{%
	\let\@ifnextchar\new@ifnextchar
	\left\lbrace
	\def\arraystretch{#1}%
	\array{@{}l@{\quad}l@{}}%
}
\makeatother
\extrafloats{100}

\begin{document}

\title{Quantum Speed-Up at Zero Temperature via Coherent Catalysis}

\author{Gabriel A. Durkin} \email{gabriel.durkin@uber.com}
\affiliation{Uber Technologies, Inc., 1455 Market St, Suite 400, San Francisco, CA 94103}
\affiliation{Berkeley Center for Quantum Information and Computation, University of California, Berkeley, CA 94720}

\date{\today}

\begin{abstract}
It is known that secondary non-stoquastic drivers may offer speed-ups or catalysis in some models of adiabatic quantum computation accompanying the more typical transverse field driver. Their combined intent is to raze potential barriers to zero during adiabatic evolution from a false vacuum to a true minimum; first order phase transitions are softened into second order transitions. We move beyond mean-field analysis to a fully quantum model of a spin ensemble undergoing adiabatic evolution in which the spins are mapped to a variable mass particle in a continuous one-dimensional potential. We demonstrate the necessary criteria for enhanced mobility or `speed-up' across potential barriers is actually a quantum form of the Rayleigh criterion. Quantum catalysis is exhibited in models where previously thought not possible, when barriers cannot be eliminated. For the $3$-spin model with secondary anti-ferromagnetic driver, catalysed time complexity scales between linear and quadratically with the number of qubits.  As a corollary, we identify a useful resonance criterion for quantum phase transition that differs from the classical one, but converges on it, in the thermodynamic limit.

\end{abstract}

\maketitle

\section{Introduction}
In computer science, computational tasks may be crudely divided into two categories: easy and hard. Easy problems are soluble in a time limit $t^{*}$ that scales polynomially with $n$, the number of available  computing resources (bits or qubits); $t^* \sim n^{\alpha}$, whereas hard tasks might take an exponentially long time to complete; $t^* \sim\alpha^{n}$. Much of the interest in \emph{quantum} computing has been fueled by the possibility that classically hard problems can sometimes become `easy' when performed by a particular quantum algorithm running on a quantum computer.

\begin{figure}[t!]
	\includegraphics[width=3.2in]{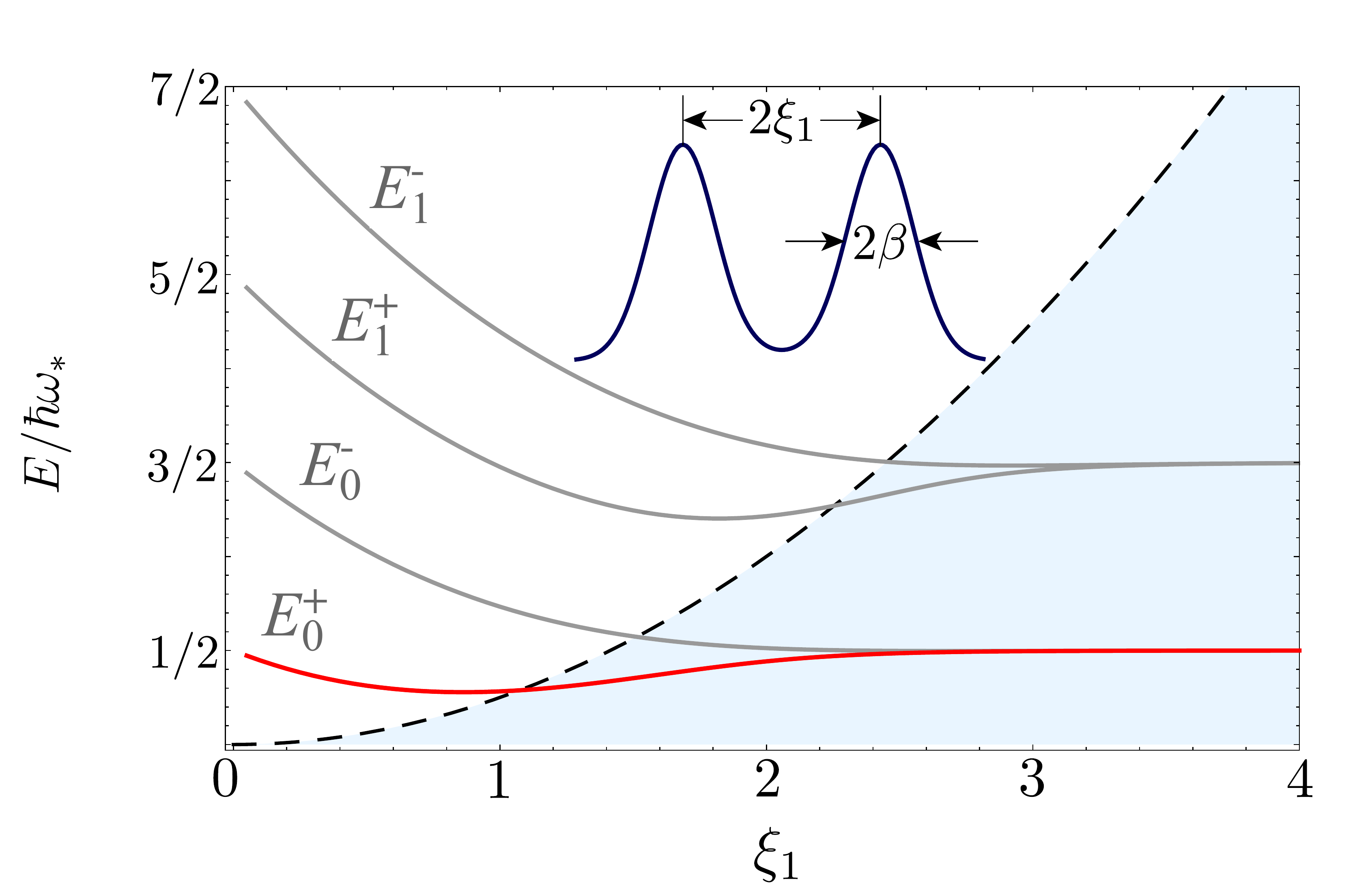}
	\caption{\label{pairing} Pairing of energies during transition from unimodal to bimodal ground state. Here $\xi_{1}$ is a displacement variable and $\beta$ a `width' variable. This important transition we call the quantum Rayleigh limit: $\xi_1 \sim \beta$, derived later in eqn.\eqref{Rayleigh}:
		As plotted, adjacent even and odd eigenvalues $E^{\pm}_k$ of the piecewise-parabolic potential are seen to `pair up', to become doubly degenerate, as the scaled separation of the wells increases beyond the order of the well widths, i.e. $\xi_1 \gg \beta$ (above for the case $\beta = 1$). In this limit (blue shaded region) the energies are well below the barrier height (dashed line) and the intra-pair spectral gap decreases exponentially with the barrier height $V_0$. A sketch of the ground-state wavefunction is shown (upper middle) in the limit of well-separated wells. For  $\xi_1 < \beta$, violating the Rayleigh criterion, the wells coalesce and the vacuum energy $E_0^{+}$ (red) and other eigen-energies subsume the barrier. At the leftmost edge the two wells merge completely, and eigenvalues revert to those of a simple harmonic oscillator: $(n+1/2) \hbar \omega$.}
\end{figure}

%As another example, the lower plot show the same phenomenon occurring in the isotropic Lipkin-Meshkov-Glick (LMG) model\cite{lipkin1965validity},  a uniformly coupled Ising spin system that maps onto a particle in a double-well potential. In the lower plot, $2 z_1$ is the separation of the well minima in the coordinate $z = m/j \in [-1,1]$ where $j$ and $m$ are total and magnetic spin quantum numbers, respectively.

One quantum computing paradigm particularly well-suited to the solution of optimization problems is  adiabatic quantum computing\cite{kadowaki1998quantum,farhi2000quantum}. In this model, an ensemble of quantum bits (qubits) is initialized in ground state of a trivial Hamiltonian. This could be, for instance, associated with a strong linear magnetic field. During the execution of the algorithm the Hamiltonian is smoothly and continuously changed or \emph{annealed} into the `target' Hamiltonian encoding the original computational task. This ground state  represents a globally optimal solution to that task. If the quantum annealing from initial to target Hamiltonian occurs sufficiently slowly or adiabatically, the system  remains in the instantaneous ground state throughout, guaranteeing that the optimal solution is recovered in finite time. The system always stays at the lowest point on the energy surface or `cost function' landscape. This approach differs quite dramatically from classical approaches to optimization, possibly involving gradient descent techniques, where it can be impossible to know whether a recovered solution corresponds to a local or global minimum. This also assumes the cost function in parameter space is sufficiently smooth that gradient information can be derived -- not often the case in combinatorial optimization. 

While success is guaranteed in finite time by the adiabatic theorem\cite{born1928beweis}, the energy landscape may become highly non-trivial during the annealing process; the ground state must navigate through a landscape of hills and valleys that spring up around it as the algorithm progresses. Large time penalties are suffered when energy barriers, of height $O(1)$ on the scale of the Hamiltonian, emerge between the current state of the system and the true minimum energy. These energy barriers are associated with a first order phase transition during the adiabatic process. Under ideal circumstances the algorithm is executed at zero temperature, to preserve the system in its ground state, so there is no possibility of thermal activation over the barrier in question. The only remaining option is for the quantum state to tunnel through the barrier to the true ground state on the other side, exactly as first discussed by Georg Gamow in his famous paper of 1928 describing alpha particle decay, \cite{gamow1928quantentheorie}. The ground state is seen to jump in a discontinuous way between configurations and the phase transition is described as `first-order'. This phenomenon is possibly the Achilles heel of quantum annealing: Tunneling, only possible via quantum mechanics, is also an exponentially slow process in $w$, the barrier width: $t^{*} \propto c^{w}$. Consequently, for problems which exhibit barrier widths scaling positively with system size $n$ (number ot qubits), the exponential delay in adiabatic passage at the phase transition produces an exponential slowdown in performance. This is reflected in an overall exponential-scaling time-to-solution with $n$. The problem instances  that feature such first-order phase transitions seem to belong (unavoidably) to the class of `hard' problems\cite{jorg2010energy}.

In this note, we will explore, in a fully quantum setting, techniques by which those barriers may be reduced, by the introduction of additional control fields or interactions. The idea that secondary control interactions might eliminate barriers, turning hard problems into easy ones, was proposed $16$ years ago in Ref. \onlinecite{bulatov2002total} and the idea resurfaced more recently in Ref. \onlinecite{seki2012quantum} where the secondary driver was of a specific non-stoquastic type. Numerical support for quantum speed-up in instances of spin glasses was given in Ref.\onlinecite{hormozi2017nonstoquastic}.  

Here we refer to controlled barrier suppression in a more general sense as `coherent catalysis'. This invites a comparison with classical chemical kinematics in a sense that was perhaps first used for quantum information in Ref.\onlinecite{jonathan1999entanglement} and that surfaced again recently in the context of adiabatic quantum computing in the excellent review by Albash and Lidar \cite{albash2018adiabatic}.  The application of the `catalyst' (secondary driver) during annealing  lowers the activation energy of the migration from false to true ground state. The purity of the quantum state is preserved, hence the process is coherent.

We begin with an examination of a prototype double-well system, for which we establish the necessary conditions for a crossover from exponential to polynomial time complexity. This is associated with a qualitative change in the quantum ground state from a bimodal to unimodal profile, and energy level `unpairing', as  illustrated in FIG. \ref{pairing}. An analogy is made with the Rayleigh diffraction limit of angular resolution in physical optics\cite{rayleigh1879xxxi} (Two point-like objects are considered resolved when the maximum of one image coincides with the first minimum of the other. When applied to two gaussian point-spread functions, the distance between the two maxima becomes comparable to the sum of their standard deviations.). 

Moving to composite systems,  we will see that the number $n$ of adiabatically-evolving qubits  plays a non-trivial role in the time complexity -- in some sense, quantum computers of mesoscopic scale might be better suited to certain classes of computational task, rather than holding fast to the naive idea that `more is simply better'. Even in a completely decoherence-free setting the limit of large $n$ will lead to a predominantly classical behaviour. A peculiarity in our analysis produces an effective Planck's constant $\hbar$ that varies \emph{inversely} with the number of qubits, attaining values much larger than $10^{-34}$ in systems of modest size (to be clarified in section \ref{hbar_sect}). As a result of this inverse relationship larger ensembles of qubits exhibit weaker quantum behaviour.

The novelty of our technique is to move beyond a conventional `mean-field' calculation by inclusion of phenomena derived from or modified by the zero point (vacuum) energy of the quantum system as it evolves through the shifting potential landscape. (In contrast, the mean field description reproduces only that potential energy surface and ignores kinetic energy completely.) Even at zero temperature a quantum system possesses vacuum energy and there exists the possibility that it overwhelms any adjacent barrier and/or `delocalizes'; this effect is magnified for a large effective $\hbar$.

To showcase the utility of these results, we re-examine the widely-studied quantum 3-spin model, presenting 3-body interactions of uniform strength between all qubit triples. It is revealed, contrary to previous thinking, that a crossover from hard to easy solution is indeed possible, with non-stoquastic drivers. (In the appendix we further discuss the somewhat simpler Lipkin-Meshkov-Glick model \cite{Fallieros1959,lipkin1965validity} that again has long range order but only 2-local interactions  in the presence of both transverse and longitudinal fields. Such a setting may be more amenable to near-term experimental verification of coherent catalysis, given some of the latest advances in quantum computing hardware \cite{zhang2017observation}.) We choose to examine these highly symmetric `toy' Hamiltonians with no topological features as they are analytically tractable yet exhibit first and second order phase transitions typical of real-world optimization problems.

\begin{figure}[b]
	\includegraphics[width=3.in]{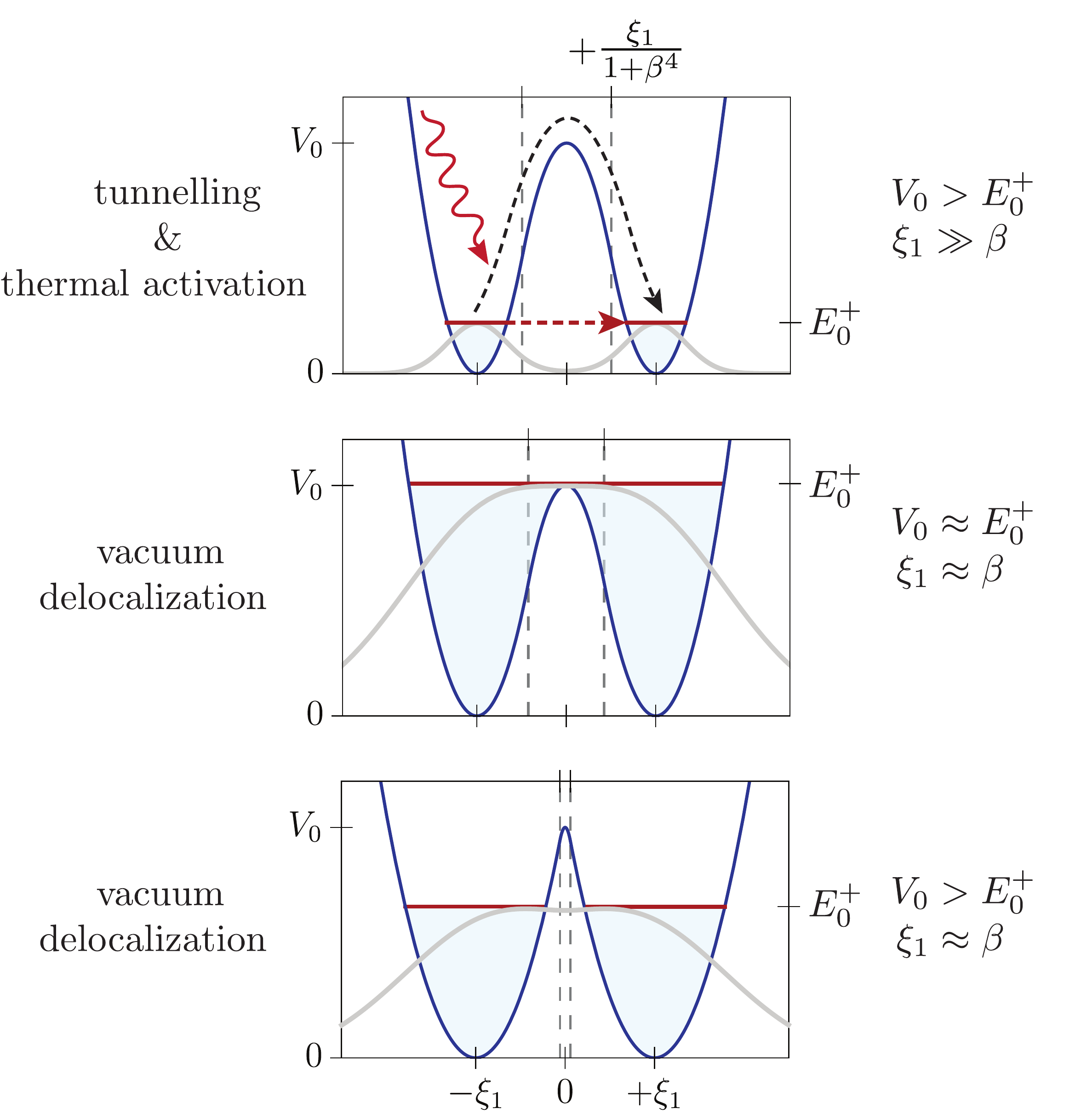}
	\caption{\label{db_well}
		Three quantum transport mechanisms exist in a double well during annealing. Potential barrier (dark blue curve) has height $V_0$ and the vacuum energy is $E_{0}^{+}$ (red horizontal line). The inverted parabola centered at the origin is stitched to two parabolic wells centered on $\pm \xi_1$  with  characteristic width $\beta$. The join location is $\xi = \pm \xi_1/(1+\beta^4)$  (vertical dashed lines in all subplots). Upper panel indicates two well-understood mechanisms of quantum tunnelling (red dashed line) through the barrier and classical activation over the barrier (black dashed line) following absorption by a thermal photon from the environment. The ground state  is shown as a grey profile in all three panels. Tunnelling and thermal activation are  possible transport mechanisms when the wells are far apart, $\xi_1 \gg \beta$, and the barrier is larger than the vacuum energy, $V_0 > E_{0}^{+}$. There is, however, another transport mechanism, \emph{vacuum delocalization}, that comes into play when the width of the ground state in one isolated well approaches the well-separation, i.e. $\beta \approx \xi_1$, perhaps initiated by some external catalysis. This 'Rayleigh limit' may occur when the vacuum energy subsumes the barrier, as in the middle panel. While sufficient for  delocalization, this is by no means a necessary condition -- as the lower panel demonstrates. The  potential in this lower panel has values $( \alpha, \beta) \mapsto (2, 2)$ , as compared with $ \mapsto (3,1)$ in the upper panel,  and $\mapsto (1.08,1.08)$ in the middle panel. }
\end{figure}
%Both $\xi = z/\sigma_{*}$ and $\beta = \sigma_1/\sigma_{*}$ are measured in units of %$\sigma_{*}$, the characteristic width at the barrier summit. (See appendix A)
\section{Introducing Quantum Transport by  Vacuum Delocalization}

It is said that `a rising tide raises all ships'. Traditionally one examines the potential landscape of  quantum annealing problems in isolation, seeking insight from the landscape's shifting topology as the annealing progresses. That level of analysis, however, may miss some subtleties and features that allow quantum speed-ups where they were previously thought not possible.  In essence, the Hamiltonian has both potential and kinetic energy, and the latter may play a significant role in transport. The evolution of the ground state components is not that of a classical hill-walker exploring the contours of the potential landscape, nor that of a quantum particle tunneling underneath the barrier -- it is more akin to a ship buoyed up over it \emph{on the sea of its own vacuum energy}, FIG. \ref{db_well}.

\begin{figure}
	\includegraphics[width=3.in]{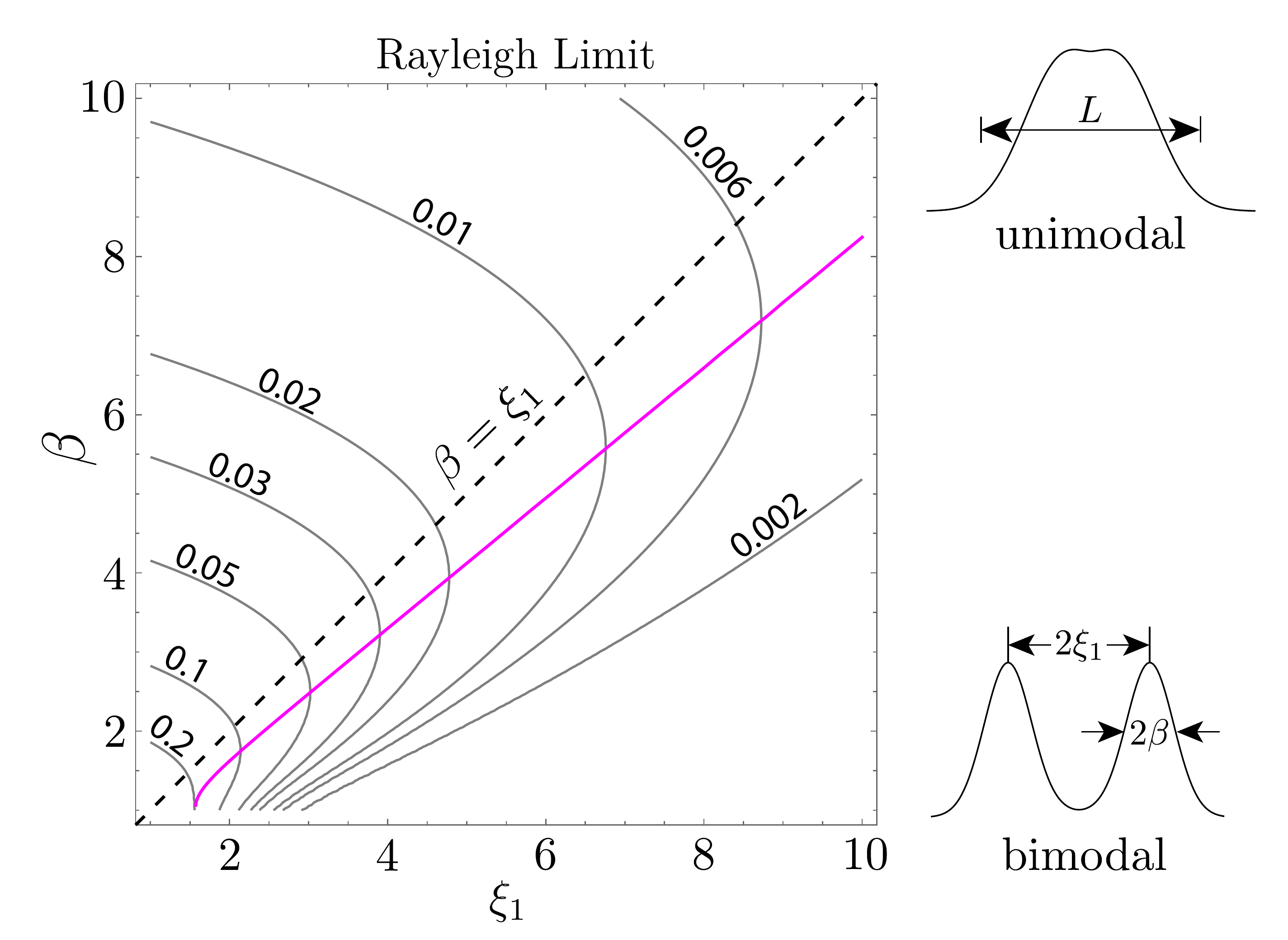}
	\caption{\label{rayleigh}
		Contour plot of iso-gaps (constant spectral gap $\Delta_{01}$) in $\{ \xi_1,\beta\}$ parameter space for the potential of section \ref{double_well}. Rayleigh limit is indicated by magenta line.  As in previous figure, effective width of the wells vs. the width of the top of the barrier is characterized by ratio $\beta = \sigma_1/\sigma_{*}$ from eqns.\eqref{rescaled}. The distance between the well minima is $2 \xi_1$ in the same scale-free units.  Gap $\Delta_{01}$ is exponentially small when the ground state is split into two localized components (lower right sketch), one in each well; $\xi_1 > \beta$. If these two parameters become comparable the gap approaches $\hbar \omega_1$  as  the wells effectively merge (upper right sketch). For a bulk unimodal state of effective width $L$ the energies and gaps will scale $\propto L^{-2}$ (particle in a box). For fixed well separation $\xi_1$ the magenta curve traces $\beta$ values associated with a maximum gap, i.e. where the indicated iso-gap contours are vertical. This line asymptotes to $\beta \sim 0.82 \xi_1$. For larger $\beta$ (above the magenta line)  $\Delta_{01}$ remains large but the well energies and gaps decrease as $1/\beta^2$, no longer exponentially.}
\end{figure}

 It has been discovered in certain quantum annealing models (by an external control field or coupling) the potential landscape of complex hills and valleys may be altered in the proximity of a phase change, when the quantum state tunnels from one potential well to another through an intervening barrier. The wells on either side of the barrier begin to coalesce as  the intervening barrier is suppressed, `softening' the phase change from first order to second order (or discontinuous to continuous). It is sometimes assumed that the barrier must be  completely razed for such a qualitative change in the characteristics of the phase change to occur. It is our observation that lowering the barrier to the scale of the vacuum energy is sufficient. This allows \emph{the vacuum state to subsume the barrier and delocalize}. In tandem, the adiabatic transfer of the quantum state between the wells proceeds at an exponentially increased rate. Interestingly, we shall see that the more fundamental condition for this enhanced mobility is that the ground state profile be at the point of coalescing from bimodal to unimodal, see FIG. \ref{db_well}. This limit we refer to as the Rayleigh limit, for obvious reasons, examined further in FIG. \ref{rayleigh}. At such a point, the system may be considered as a `particle in a box', the dimension of the box corresponding to the width of the unimodal ground state at the phase transition.  (In the antithetical scenario a large intervening barrier greatly exceeds the vacuum energy, permitting only exponentially-slow quantum tunneling of a localized state from one side to the other.)
 
 First, a thorough investigation of a double well will provide a more mathematical underpinning to the above remarks.

\section{Quantum Transport in a Double Well} \label{double_well}
Since the performance of the adiabatic algorithm stems from the ability of the ground state to conquer potential barriers as it evolves towards the target state, let us examine this simplest of possible scenarios --  that of a generic symmetric double well in one dimension, with real coordinate $z$, and centered on $z=0$. Such models are used extensively with great success to explain  phenomena such as diatomic molecular bonding in chemistry and microwave frequency oscillations between the vibrational modes of the ammonia molecule. 

 As the distance between two wells increases, the spectrum of energy levels is observed to pair up in doublets, each containing adjacent orthogonal symmetric and anti-symmetric eigenstates. Doublets are separated in energy by $\sim \hbar \omega$ (the characteristic energy of one well taken in isolation) but the intra-doublet splitting itself shrinks exponentially small in the well separation. See FIG.\ref{pairing}. By this exercise, and in the spirit of Gamow's work \cite{gamow1928quantentheorie} mentioned earlier, we wish to gain insights about  any crossover in the scaling behaviour of these spectral gaps as the wells distance varies. For the execution of the adiabatic algorithm, the gap size fundamentally dictates time to solution. To foster confidence in the veracity of this statement we present the following illustration. 
 
  Supposing  the system begins in the ground state of the left well, we can calculate the probability it will migrate to the right well. Assume the true ground state of the system is a real-valued positive set of amplitudes that corresponds to an equal superposition $\psi^{+}_0(z) = (\psi_0(z-z_1) + \psi_0(z+z_1))/\sqrt{2}$ of the local simple harmonic oscillator ground states confined to the left and right wells. This is a reasonable assumption for wells  separated by a wide barrier. Near a minimum the potential is by definition quadratic: $V(z \approx z_1) = V_0 + V''_0 (z-z_1)^2$, we recall the eigenstates of a simple harmonic oscillator are Hermite functions, with the ground state Gaussian-distributed: $\psi_0(z) \propto \exp \{-z^2/(2 \sigma^2)\}$ where $\sigma = \sqrt{\hbar/(m \omega)}$ provides a natural length-scale and $\omega = \sqrt{V''_0/m}$ relates the energy scale of local quanta to the well-curvature. Equally, we can assume the first excited state is the anti-symmetric superposition $\psi^{-}_0(z) = [\psi_0(z-z_1) - \psi_0(z+z_1)]/\sqrt{2}$, as its orthogonality to the ground state is guaranteed, even when the overlap $\int \psi_0(z-z_1) \psi_0(z+z_1) d z$ becomes substantial. The symmetry of the potential energy $V(z) = V(-z)$ guarantees that eigenstates will have definite even/odd parity, $\psi^{\pm}_k(z) = \pm \psi^{\pm}_k(-z)$ where $k$ labels a particular doublet. The Sturm-Liouville theorem\cite{zettl2005sturm} dictates that in one dimension there can be no degeneracies, and that the odd/even doublets will be paired with the odd states above the even ones in energy\footnote{An additional assumption is that the energy scales of the doublets versus the inter-doublet spacing is large enough that confining analysis to the two-state subspace is justified. This assumption becomes invalid for wells that are closer together, the overlap between the localized oscillators becomes appreciable, the intra-doublet and inter-doublet level spacing become comparable, as on the left side of FIG. \ref{pairing}. Then a different argument will be needed to relate the spectral gap to the rate of migration across the barrier. The orthonormal basis of eigenfunctions is no longer composed of symmetrized gaussians, but parabolic cylinder functions with real and imaginary arguments.}.

Now to relate the spectral gap size to the rate of migration across the barrier for the two state model: Taking the Hamiltonian to be quasi-static during the transition, a particle localized in the left well, at $t=0$ will be  $\psi_0(z-z_1) \propto  \psi^{+}_0 - \psi^{-}_0$, which evolves to:
\begin{align}
\psi(z,t) &= \exp  \{-i E^{+}_0 t /\hbar\}\psi^{+}_0  -  \exp\{-i E^{-}_0 t /\hbar\} \psi^{-}_0 \nonumber \\
&  = \exp \left\{\frac{-i (E^{+}_0  +E^{-}_0) t }{2\hbar}\right\} \nonumber \\ 
& \times\left \{ (\psi^{+}_0 + \psi^{-}_0) \cos \frac{t \Delta }{2\hbar}  +i  (\psi^{+}_0 - \psi^{-}_0)  \sin  \frac{t \Delta }{2\hbar} \right\}
\end{align}
 (ignoring normalization) and $\Delta = E^{-}_0 - E^{+}_0 $ is the energy gap. The system oscillates back and forth between the two wells at a frequency $\Delta/(2\hbar)$ and a characteristic time-scale for the migration from left to right well is  
\begin{equation}
\tau \sim h/\Delta .
\end{equation}

Armed with the knowledge that barrier migration occurs on time scales varying inversely with the size of the spectral gap, one is motivated to understand how the latter varies with the height of the barrier and the distance between the two wells. This has been a heavily researched topic in the limit of tall barriers and large separations \cite{garg2000tunnel}, however the behaviour in transition to lower barriers and small separations is not well-documented.

If we demand the double well to be continuous, smooth and piecewise-parabolic, the barrier height can be calculated:
\begin{equation}\label{barrier-height}
V_0 = \frac{m z_1^2}{2}  \left( \frac{1}{\omega_{*}^2} + \frac{1}{\omega_1^2}  \right)^{-1}.
\end{equation} 
At the barrier summit we define a characteristic frequency $\omega_{*} = \sqrt{-V''_0/m_{*}}$ and a length scale by  $\sigma_{*} = \sqrt{\hbar/ (m_{*} \omega_{*})}$;  the width of a localized ground state, were the maximum inverted. Analogously we can define $  \sigma_1 = \sqrt{\hbar/ (m_1 \omega_1)} $, the width of a localized ground state in one of the two wells with frequency, $\omega_1 = \sqrt{V''_1/m_1}$.
This `Frankensteined' potential represents a  restricted subset of all possible double wells but its modest formulation, we hope, will produce insights that generalize well to the wider domain (for instance, when we examine the quantum $3$ spin model in section \ref{3spin}.)   One might arbitrarily imagine wells with more elaborate  structure away from the well extrema, without revealing much about quantum transport mechanisms in general. As was stated earlier, it is not the barrier height, or even its size relative to the ground state energy that establishes the spectral gap and computational complexity. The more correct question one should ask is whether the distinct well components have coalesced, i.e. whether a quantum Rayleigh `resolution limit' is reached.

\begin{figure}
	\includegraphics[width=2.9in]{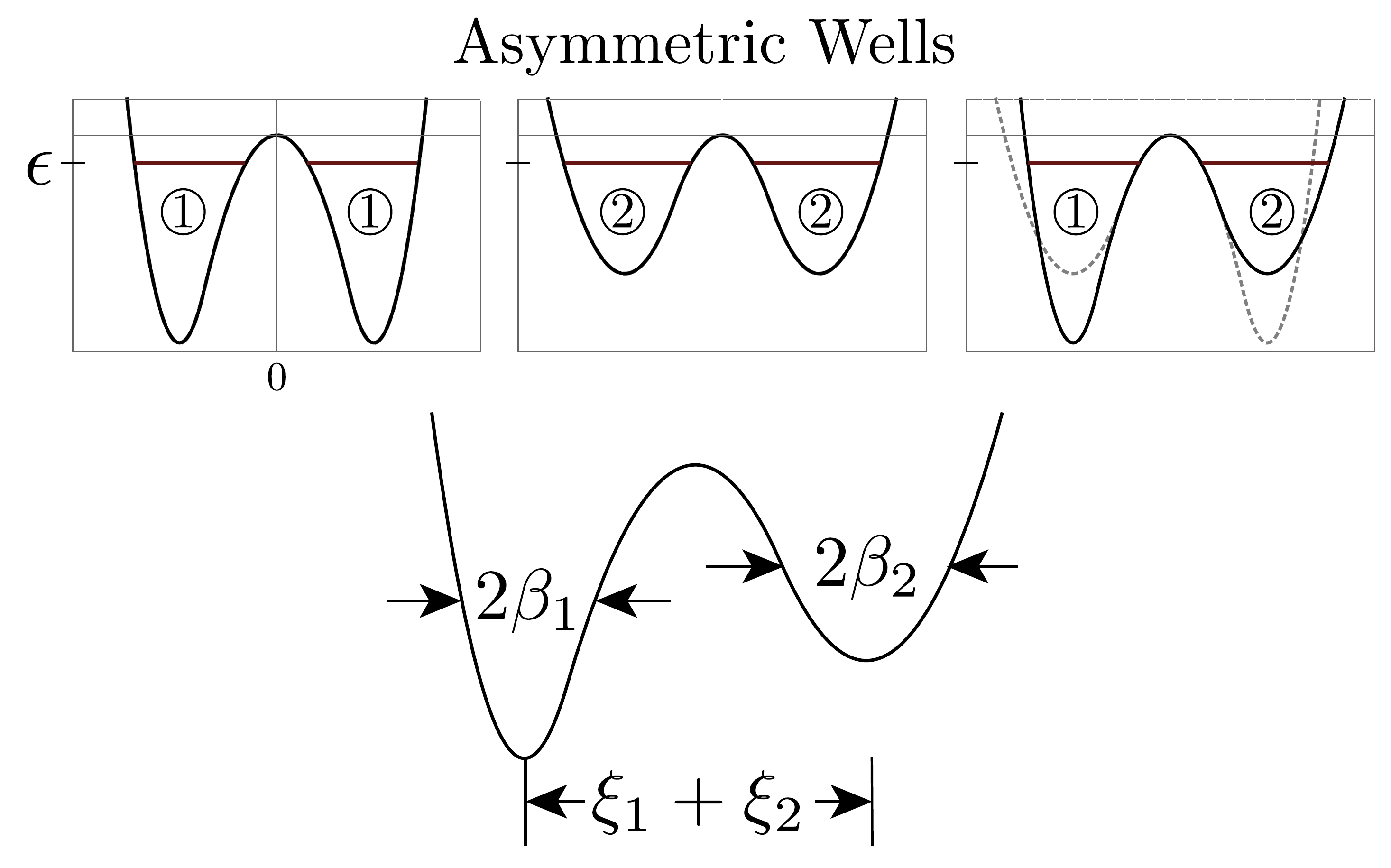}
	\caption{\label{antirayleigh} An asymmetric well may also be modelled as piecewise parabolic, with two location and two width parameters, $\{\xi_{1,2}, \beta_{1,2}\}$ respectively. Here we have chosen different width parameters $\{ \beta_1,\beta_2 \} \mapsto \{ 1, 3^{-1/4} \}$ for all plots, and equal location parameters $\xi_1 = \xi_2 \approx 1.25$. These values correspond to the optimal catalysis of the quantum $3$-spin, see eqn. \eqref{rayleigh_answer}.  To identify the point of phase transition at the barrier summit the vacuum energy $\epsilon$ of all three potentials must coincide; the asymmetric potential and \emph{both} symmetric wells from which it is composed (vacuum energy level depicted by red horizontal lines). Near the barrier summit the potential is an inverted parabola; the ground state has identical analytical form in all three cases (upper plots); a standing wave formed from right and left-moving parabolic cylinder functions, see Appendix \ref{appendix_potential} for further discussion. Without this resonance condition we cannot assume that the ground state represents a system at the point of phase transition. FIG.\ref{no_rayleigh} illustrates the different phenomenon of resonance tunneling for slow phase transitions where neither well approaches the Rayleigh limit and the barrier greatly exceeds the vacuum energy.}
\end{figure}

Detailed examination of the piecewise potential and its analytical eigenstates is presented in the appendix \ref{appendix_potential}. Eigenstates $\phi^{\pm}$ are composed of parabolic cylinder functions, e.g. the ground state $\phi^{+}$ is a superposition of such cylinder functions\cite{connor1968analytical} called a Kummer function,  eqn.\eqref{invertedGS}.

The following useful expression is derived in appendix \ref{appendix_potential} for the spectral gap as a scale-free ratio:
\begin{equation}\label{gapeqn2}
\frac{\Delta_{01}}{\hbar \omega_{*}}  = \frac{\phi^{+}_0(0) \frac{d \phi^{-}_0}{d \xi}(0) }{2 \int_{0}^{\infty} \phi^{+}_0 \phi^{-}_0 d \xi}
\end{equation}
where the denominator is the semi-overlap of the ground and excited states, and the variable $\xi = z/\sigma_{*}$ is the displacement variable (measured in units of $\sigma_{*}$, the effective width of the barrier summit).   Substituting the analytical forms of the eigenstates, the gap function of eqn.\eqref{gapeqn2} is maximized for well-separation $\xi_1 = z_1 /\sigma_{*} \mapsto 0$, demonstrating that the largest gap occurs when two wells merge. Introducing the scaled well-width $\beta = \sigma_1/ \sigma_{*}$, FIG.\ref{rayleigh} indicates a maximum of the spectral gap in $\beta$ near $\beta \approx 0.82 \: \xi_1$. For distinct, localized wells ($\xi_1 \gg \beta $) the gap quickly vanishes to become exponentially small. For fixed $\xi_{1}$ and larger $\beta \gg \xi_1$ beyond its maximum, the gap shrinks again, but only polynomially quickly. This is because the coalesced well is now becoming wider as $\beta$ increases. (Further reduction in the barrier height occurs but is now irrelevant.) The gap decays as the inverse square, $\Delta/ (\hbar \omega_{*})  = 1/\beta^2 = \omega_1 / \omega_{*}$, as should be expected --the wells are merging and the gap maps onto that of a single harmonic oscillator: $\hbar \omega_1$.  We shall learn that the spread/confinement of the (unimodal) ground state is the harbinger of quantum mobility, rather than barrier suppression. 

In terms of the original variables,  fast adiabatic transport across a potential barrier does not depend directly on the potential barrier size, nor the curvature at the summit. Rather it depends on the violation of a Rayleigh separability criterion:

\begin{equation}\label{Rayleigh}
 z_1 > \sigma_1 = \sqrt{\hbar}\left[\frac{1}{m_1 V''(z_1) }\right]^{1/4} \; \text{(Rayleigh limit)}
\end{equation}
This criterion is entirely defined in terms of the potential curvature and coordinate extent  in the vicinity of the well minima at $z = \pm z_1$. Quantum mechanics only enters via the coupling constant $\sqrt{\hbar}$ --  its value dictates the spread or confinement of the ground state in a potential minimum. This may seem an odd remark, but it is relevant for later sections where an effective $\hbar(n)$ emerges that will depend on the number of qubits $n$, see eqn. \eqref{hbar_n}.

It is not obvious how universal such a criterion might be, and whether it might be extrapolated to wells of different shape and symmetry.  In a later section we tackle the quantum 3-spin, where we will encounter transitions between asymmetric double wells of a quantum particle with position-dependent mass.

\section{Asymmetric Potential and Resonance Condition} \label{resonance_section}

The asymmetric double-well with a high barrier and widely separated minima $\xi_{1}$ and $\xi_2$ can be characterised as being far from the Rayleigh limit: $\xi_i \gg \beta_i$. Such a scenario is less interesting in the current context of vacuum delocalization, and has  already been carefully examined, including the phenomenon of resonant tunneling, in Refs. \onlinecite{halataei2017tunnel,vybornyi2014tunnel,rastelli2012semiclassical}.

When the wells begin to coalesce and $\beta_i \sim  \xi_i$ one might ask now what are the conditions for quantum catalysis when the potential has asymmetry? To begin with, we must consider whether the system is at the point of a quantum phase transition. In the symmetric well it was guaranteed that the ground state represents a phase transition due to its inherent symmetry. But for a asymmetric double-well, the ground state may be largely confined to the wider/deeper minimum. 

To correctly identify the point of transition we should look to the barrier summit. A key observation is that the asymmetric piecewise parabolic potential still has a maximum of unit curvature there. Locally this inverted parabola has travelling wave solutions $\psi(\pm z)$; parabolic cylinder functions with imaginary arguments\cite{connor1968analytical}, moving to the left and right, see appendix A. (Bound states of double well must share the same eigenvalue and the travelling waves are mirror images of each other.) The overall solution in the vicinity of the barrier is a linear combination of these, $A \psi(z) + B \psi(-z)$. 

Given the context of phase transition, if we associate the minimum spectral gap with the maximum ground state variance, this maximum is only possible when the state is as equally distributed as possible between the two asymmetric wells. By matching energy levels and introducing the boundary condition at the barrier summit $\psi'(z_{*})=0$, it's as if we have introduced a double-sided mirror -- each side of the barrier has a ground state that is one half of a fully-symmetric double-well. This is the key to an evenly distributed wavefunction between the two asymmetric wells. So the real-valued symmetric combination ($A=B$ above) is the only possibility. 

Once this Kummer wavefunction \eqref{invertedGS} is matched to the solutions further out in each of the two wells of different widths and depths, it will be stitched to a different decaying solution as $z \mapsto \pm  \infty$. The join points will differ on each side because of the asymmetry, $\xi = -\xi_1/(1+\beta_1^4)$ and $\xi = +\xi_2/(1+\beta_2^4)$. The overall ground state will look like the ground state of a symmetric double-well for $z<0$ joined at $z=0$ to the ground state for a \emph{different} symmetric double-well for $z>0$. It is necessary that the two symmetric wells have the same ground state energy; all the stitched together ground state components must share the same eigenvalue to represent a composite eigenstate. See FIG.\ref{antirayleigh}.

Thus we have a \emph{resonance} condition analogous to the one for tunneling through high barriers mentioned at the start of this section: Phase transitions in asymmetric double-wells occur when the ground states of the two symmetric double-wells (from which the asymmetric well is derived) have the same energy deficit below the barrier summit.
\begin{figure}
	\includegraphics[width=3.in]{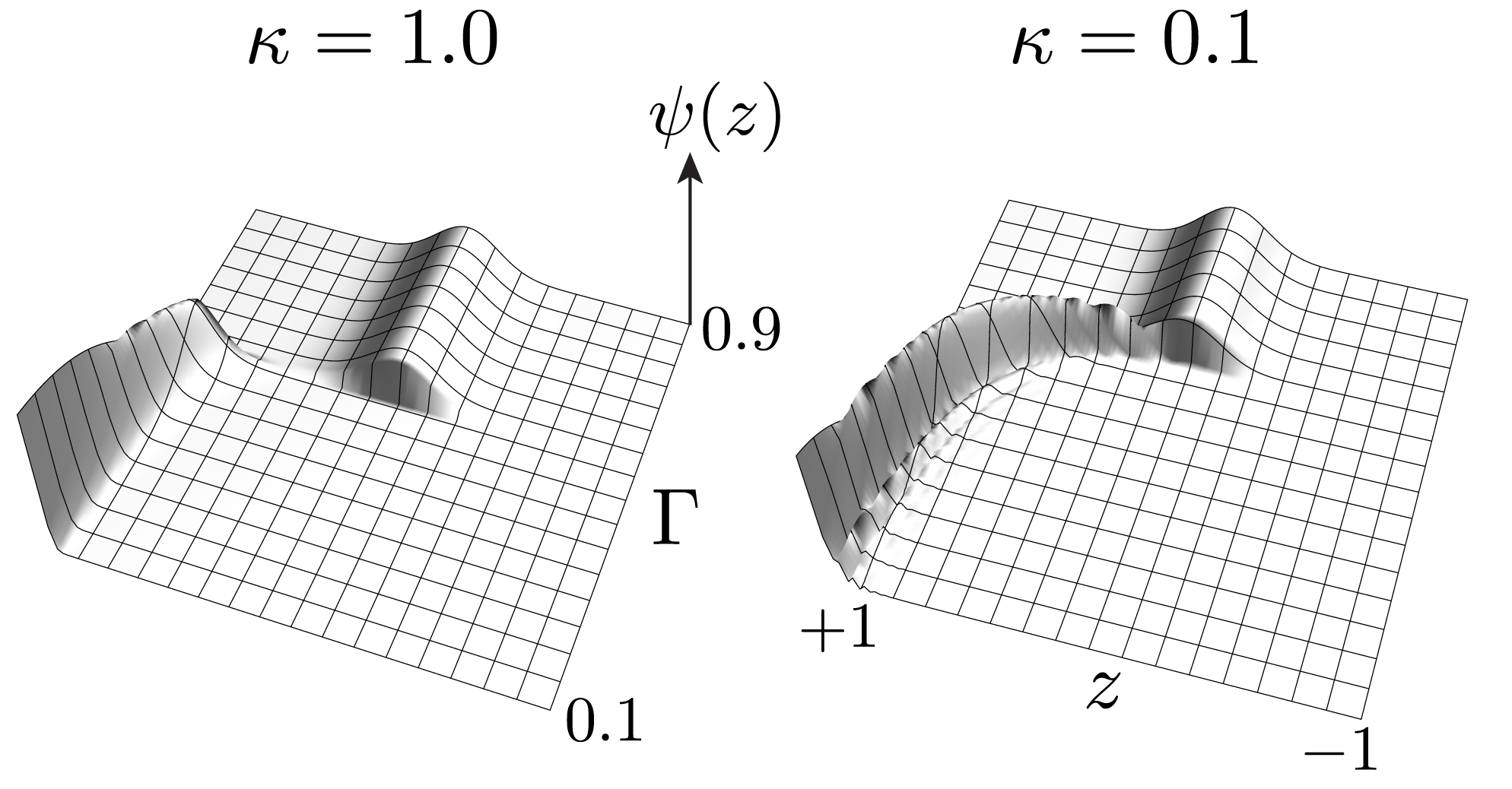}
	\caption{\label{kappa_effect}
		In the quantum 3-spin model, increasing the anti-ferromagnetic coupling governed by $(1-\kappa)$ from eqn.\eqref{3spinwithkappa} `softens' the discontinuous or first-order phase transition occurring for the external magnetic  field parameter $\Gamma = 0.565$, to approach a continuous or second-order transition nearby.  This is illustrated above for $n=100$ spins. For $\kappa = 1$ the discontinuity in the ground state (amplitudes $\psi(z)$ on the vertical axis) is quite apparent as it tunnels from the paramagnetic phase centered on $z = m/j = 0$ to the ferromagnetic phase near $z = +1$. For $\kappa = 0.1$,  this transition has been `smeared out' by the contribution of the non-stoquastic driver $+\hat{J}_{x}^2$; the state distribution changes continuously with $\Gamma$ in the second plot. }
\end{figure}
	
\section{introducing the Ferromagnetic 3-Spin Model} \label{3spin}

Now let us apply this delocalization transport mechanism to systems of $n$ spins or qubits. We can define a quantum annealing Hamlitonian as follows:
\begin{equation}
\hat{H} = - \Gamma \frac{\hat{J}_x}{j} - (1-\Gamma) \frac{\hat{J}_{z}^{p}}{j^p} 
\end{equation}
where $\Gamma$ is an annealing parameter. The operators $\hat{J}_{x,y,z}$ are associated with  spin along the three Euclidean axes. Scalar $j$ is a total spin quantum number, and $p$ is an integer power. The control parameter $\Gamma$ is typically initialized at $1$ and reduced slowly and smoothly to $0$. In the current context $\Gamma$ is associated with the strength of a transverse magnetic field (along the $x$ direction). In the usual sense, $\hat{J}_{z}^{p}$ is therefore the `target' or `problem' Hamiltonian. (The quantum annealing prepares the ground state of a problem Hamiltonian.)

\begin{figure*}
	\includegraphics[width=6.in]{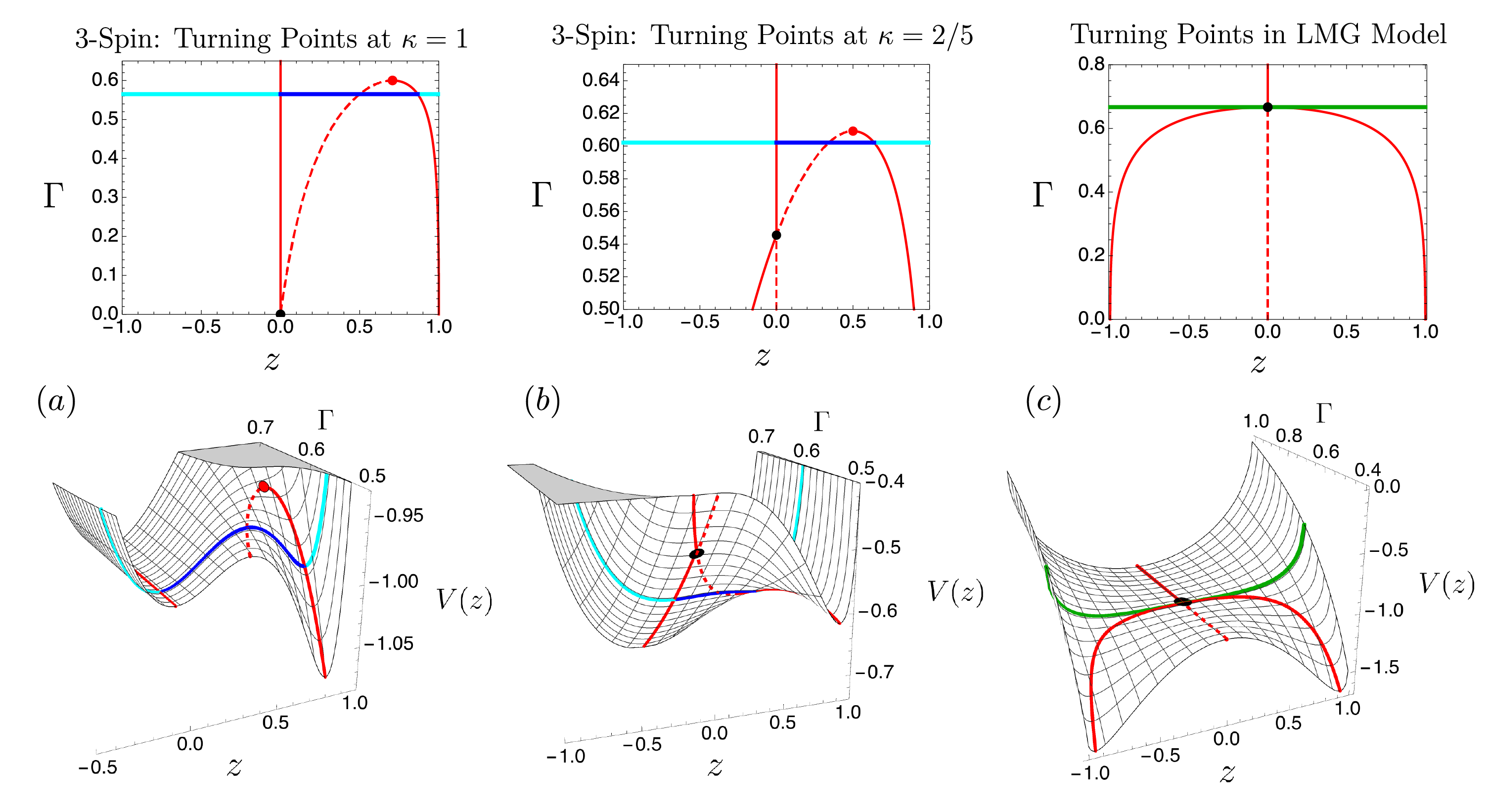}
	\caption{\label{PT}
		Zero temperature energy surfaces $V(z)$  associated with quantum annealing as a transverse field  $\Gamma$ is lowered adiabatically (upper row of 2D plots provide an overhead plan view): Coordinate $z = m/j \in [-1,1]$ is the ratio of the magnetic quantum number $m$ to total spin $j$. For an $n$-qubit ensemble confined to the fully symmetric subspace, the total spin $j = n/2$.  Loci of the maximum (dashed red line) and minima (unbroken red line) are indicated, as are births of new maximum/minimum pairs (red dots), and second order phase transitions where a minimum becomes a maximum (black dots). These latter points are associated with lines of zero curvature, $V''(z)=0$. As $\Gamma$ is reduced, the ground state evolves in (a) and (b) by a first order phase transition connecting $z=0$ via tunneling through an intervening barrier ( dark blue contour) to reach the global minimum at $z=1$. The blue/cyan contours trace out the potential wells when they have equal depth, as defines the classical first order transition. (In the 2D plan view, the blue transition line intersects the red dashed line associated with a potential maximum, a signature of barrier penetration.) To contrast, case (c) illustrates the  Lipkin-Meshkov-Glick (LMG) model \cite{Fallieros1959,lipkin1965validity}, in which the ground state smoothly evolves from being localised at $z=0$ to $z= \pm 1$, bifurcating continuously near the critical point $\Gamma_c = 2/3$  (green line). The `gentler' phase transition here is described as second order or continuous; the green contour intersects no intervening $V(z)$ maximum, there is no barrier penetration.  Cases (a) and (b) present the p-spin model for $p=3$ which, at least classically, presents an unavoidable first order phase transition; the barrier cannot be fully suppressed to zero. Case (b) shows the modified potential produced by the presence of the non-stoquastic catalyst term in the p-spin Hamiltonian, a transverse ferromagnetic  coupling of strength $(1-\kappa)$, as described in section \ref{3spin}. The catalysis softens the transition, bringing the first and second order transitions points in close proximity (red/black dots of top middle figure) and effecting \emph{partial} barrier suppression.}
\end{figure*}

As it is written, the Hamiltonian is bounded: $|\hat{H}| <1$ This model is called quantum ferromagnetic $p$-spin, and we will explore the $p=3$ case. It is worthy of a comment that the limiting case $p \sim \infty$ faithfully represents the annealing formulation of Grover's unstructured search problem, which belongs to the class of hard problems mentioned in the introduction. Since, for example, $\hat{J}_{z} = \sum_i \sigma_{z}^{(i)}/2$ is a collective spin operator, the interaction term with $p=3$ gives rise to $3$-local interactions of type $\sigma_z^{(i)} \otimes \sigma_z^{(j)} \otimes \sigma_z^{(k)}$. This may seem at first to be somewhat unphysical. The advantage for us is that  despite the uniform and infinite range couplings and no topological features the model exhibits a first order phase transition for a particular $\Gamma_c$. It may provide insights about the nature of `hard' problems in quantum annealing.

By the addition of a second control parameter $\kappa \in [0,1]$, that varies the strength of a transverse anti-ferromagnetic coupling $\sigma_x^{(i)} \otimes \sigma_x^{(j)}$, the $3$-spin model becomes  more interesting:
\begin{equation}\label{3spinwithkappa}
\hat{H} = - \Gamma \frac{\hat{J}_x}{j} - \kappa (1-\Gamma)  \frac{\hat{J}_{z}^{p}}{j^p} + (1-\Gamma) (1 - \kappa) \frac{\hat{J}_{x}^{2}}{j^2}.
\end{equation}
It should be noted that this second control Hamiltonian $+ \hat{J}_{x}^{2}$ has opposite sign to the other terms which implies that in the computational basis the off-diagonal terms are no longer real and non-positive. This is a definition of non-stoquasticity. It has been conjectured that the inclusion of such non-stoquastic terms might be crucial to any speed-up of quantum annealing over classical computation. Indeed, non-stoquastic Hamiltonians may not be simulated efficiently by classical algorithms. Seki and Nishimori proved in Ref.\onlinecite{seki2012quantum} that for p-spin models of $p \geq 4$ the inclusion of the non-stoquastic term above can, during the annealing schedule, circumvent the first order phase transition. For $p =3$, the mean-field analysis they performed indicates the first order phase transition should persist, resulting in an exponential slowdown of the adiabatic evolution. We will show that actually this is not the case; even for $p=3$ there is the possibility of non-stoquastic speed-up.

The non-stoquastic term has the effect of widening the spectral gap at the phase transition. Equivalently, the free energy landscape is altered such that potential barriers are suppressed entirely into second order phase transitions. Traversing the fully lowered barrier, the ground state no longer jumps discontinuously at the phase transition; rather it stretches across the valley floor to occupy the other well, with amplitudes that `smear' across the intervening coordinate space, see FIG. \ref{kappa_effect}.

\section{Mean Field Picture}

In a conventional treatment one proceeds with a mean field analysis. In such an approach, all $n$ qubits are unentangled and identical, collectively forming a large spin coherent state:
\begin{align}
| \psi \rangle =& \cos (\theta/2) | 0 \rangle  + \sin (\theta/2) | 1 \rangle \\
| \Psi \rangle  = & | \psi \rangle^{\otimes n} 
\end{align}
This state, because all the qubits are identical, is also in the $j=n/2$ fully-symmetric subspace. Also note the qubit is confined to the $x-z$ plane just like the Hamiltonian, and that $\theta$ is the polar angle made by spin coherent state  with the $x$ axis.
If we take the expectation value of $\hat{H}$ with $| \Psi \rangle$  we can write it in terms of $\{\Gamma, \kappa, \theta\}$ parameters:
\begin{equation}
\langle \hat{H} \rangle = - \Gamma \cos \theta - \kappa (1-\Gamma) \sin^p \theta + (1-\Gamma)(1-\kappa) \cos^2 \theta
\end{equation}
In Cartesian coordinates,  we may express $\sin \theta = z$ and $\cos \theta = x = \sqrt{1-z^2} $ and introduce the annealing ratio $\gamma = \Gamma/ (1-\Gamma)$:
\begin{equation}\label{meanfield_VZ}
\frac{\langle \hat{H} \rangle}{\Gamma}  = V(z) = - \sqrt{1-z^2} -\left[\frac{ \kappa}{\gamma} \right] z^p +\left[\frac{1-\kappa}{\gamma}\right] (1- z^2)
\end{equation}

This is a mean field description of the energy $V(z)$ as a function on the line $z = m/j \in [-1,1]$ where $m$ is the magnetic quantum number. We continue in the fully symmetric space of maximum spin $j = n/2$, as the Hamiltonian always commutes with the total spin operator $\vec{J}^2 = \hat{J}_x^2 + \hat{J}_y^2 + \hat{J}_z^2$ for all $\{\Gamma, \kappa \}$ values. This energy function on the line will continuously change as $\kappa$ and $\Gamma$ are varied, and if the changes are made adiabatically, the spin configuration remains in the minimum of this function. Equivalently, the overall spin coherent state is like a macroscopic pointer oriented in the direction $\theta_0$ associated with the minimum energy.

Of course, even though much can be gained from this classical analysis, this does not provide a complete picture. The spins are highly coupled with long range order and during its evolution the system undergoes a first or second order phase transition. It is hard to believe that a description devoid of entanglement and other quantum properties will capture the correct characteristics in proximity to the phase transition where quantum features are dominant (e.g. peaks in entanglement and quantum Fisher information). Recall also that the bottlenecks occurring in this critical region dictate the overall time complexity of the algorithm. We now illustrate this shortcoming with the $p=3$ case. 

\section{Full Quantum Model of 3-Spin with Non-Stoquastic Driver} \label{hbar_sect}

First, let us make the problem \emph{fully} quantum. We could simply proceed by numerical diagonalization of the spin Hamiltonian, but this provides little insight about the problem or its characteristic features, and does not answer questions like: Why is there a phase transition? or Why should we expect universality to the problem's behaviour for different $n$? Why should we expect the problem to compute quickly or slowly? And eventually, one might imagine running out of processing power to perform the numerical computations at  large $n$. The approach we employ is to turn the discrete spin problem into a continuous variable Schr\"odinger equation for a particle in a potential. By similar techniques, we previously studied criticality as a resource for quantum metrology in the Lipkin model \cite{durkin2016asymptotically}. Using notation $\hat{J}_z | m \rangle = m | m \rangle$, designate the ground state as $| \Psi_0 \rangle =  \sum_{m  = -j}^{+j} \psi_m | m \rangle$. Let's formulate the difference equation, or recurrence relations for the Hamiltonian, featuring $\psi_m, \psi_{m \pm 1}, $ etc., which for small $q=2$ in $\hat{J}_x^q$ will not have many entries far from the leading diagonal. Forming the inner product $\frac{1}{\Gamma}\langle m | \hat{H}| \Psi_0 \rangle $:
\begin{equation*}
 =  - \frac{\kappa}{\gamma} \left( \frac{m}{j} \right)^p  + \langle m | \left\{- \left( \frac{\hat{J}_x}{j} \right)   + \left[ \frac{1 - \kappa}{\gamma} \right]\left( \frac{\hat{J}_x}{j} \right)^2 \right\} | \Psi_0 \rangle
\end{equation*}
To proceed we recall $\hat{J}_x =( \hat{J}^{+}+\hat{J}^{-})/2$ and the action of these ladder operators is 
\begin{equation*}
\hat{J}^{\pm} | m \rangle = \sqrt{j^2- m^2 +j \mp m}\:  |m \pm 1 \rangle  . 
\end{equation*} 
Operating with the Hermitian spin operators to the left on the basis states $\langle m | \leftarrow$ gives:
\begin{align}
\langle m | \left( \frac{\hat{J}_x}{j} \right) = \frac{\sqrt{j^2-m^2}}{2 j} \; \times \nonumber \\ 
\left[ \langle m-1 | \sqrt{1+ \frac{1}{j-m}}  +  \langle m+1 |  \sqrt{1+ \frac{1}{j+m}}  \: \right] .
\end{align}
Similarly, $\hat{J}_x^q$ for $q=2$ maps the $| m \rangle$ component into itself and $| m \pm 2 \rangle$. Moving to pseudo-continuous coordinate $z = m/j  \in [-1,1]$ we introduce a small parameter: 
\begin{equation}
\hbar = 1/j  \label{hbar_n} \; \; \; \; \text{(effective Planck const)}
\end{equation} 
and rewrite $\langle m\pm 1| \Psi_0 \rangle = \psi_{m \pm 1}  \mapsto \psi(z \pm \hbar) $. The exact result (before any approximation) is
\begin{align}
\langle m | \left( \frac{\hat{J}_x}{j} \right)  | \Psi_0 \rangle = \sqrt{1-z^2} & \left[  \psi(z-\hbar) \sqrt{1+ \frac{\hbar}{1-z}} \right. \nonumber \\  +  & \; \; \left.  \psi(z+\hbar)   \sqrt{1+ \frac{\hbar}{1+z}}  \right]
\end{align}
\begin{figure}
	\includegraphics[width=2.8in]{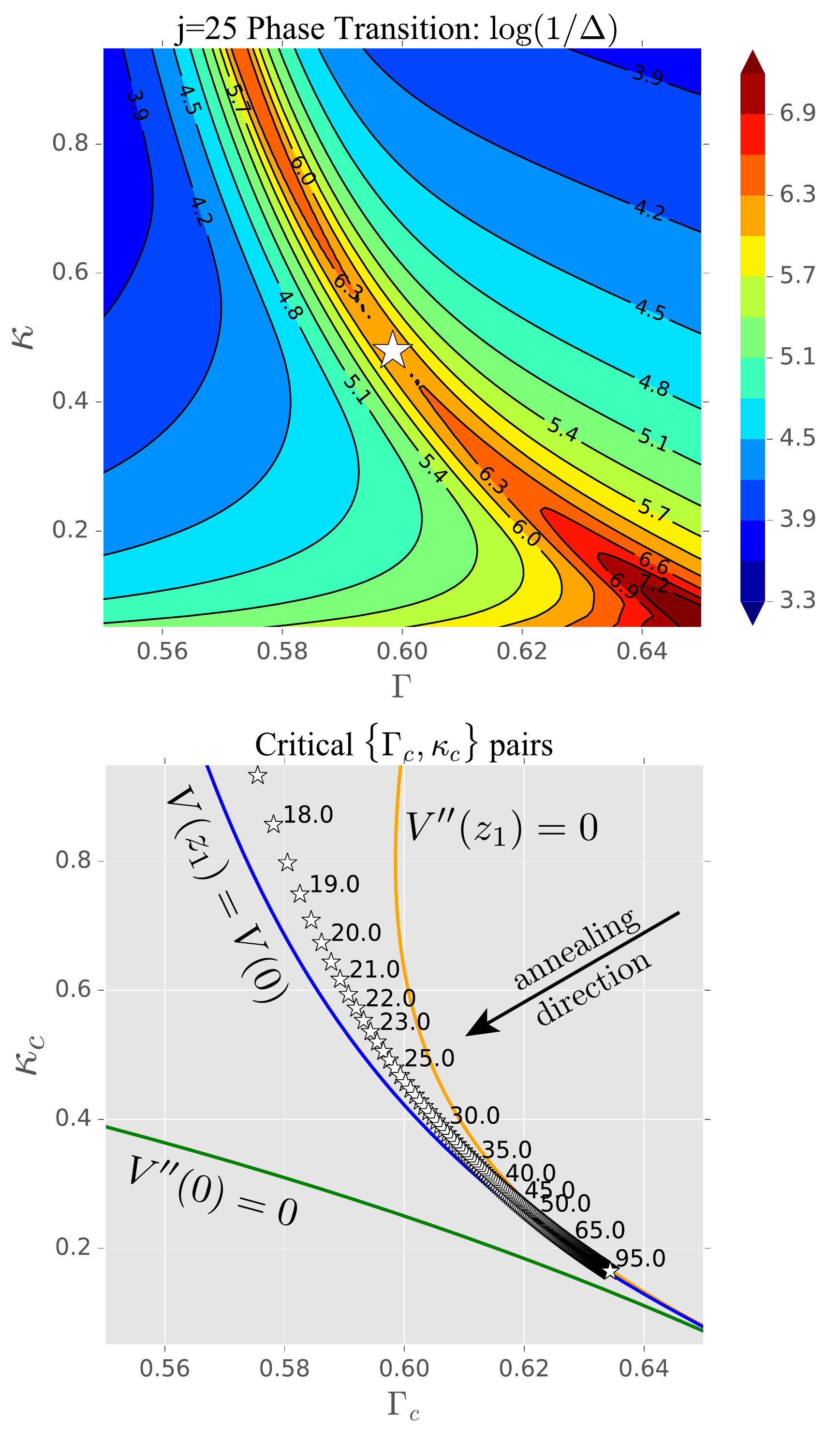}
	\caption{\label{gapstar}
		Any annealing schedule that maps $\{\Gamma, \kappa\}$ values  $\{1,1\} \mapsto \{0,1 \}$ will unavoidably traverse the minimum gap region that corresponds to the phase transition of the quantum 3-spin model, eqn. \eqref{3spinwithkappa}. The phase transition ridge may not be circumvented, as is quite apparent in the upper contour plot of the inverse gap $1 / \Delta$ for  $n = 50$ ($j=25$) . The ridge extends throughout the full parameter range $\kappa \in [0,1]$.  Near the line $\kappa=1$ the transition is first order, the associated minimum gap is always exponentially small in $j$. However, for optimized control parameters $\{\Gamma_{c}, \kappa_{c}\} = \{0.598, 0.479\}$ (denoted by a white star) the ridge defining the phase transition has a saddle,  a `maximum minimum gap'. Optimized annealing schedules that minimize computation time should pass near this point. In fact, for $\kappa \leq \kappa_{c}$ as $j \gg 1$ a polynomially-small minimum gap is always possible, even though a classical analysis of the potential landscape alone would seemingly forbid this. The lower plot shows the $\{\Gamma_{c}, \kappa_{c}\}$ parameter pairs (stars) for different labelled  $17.5 \leq j \leq 95$. (For $j \leq 17$ , optimal $\kappa_c =1$.) The  orange and blue curves describe respectively, the birth of the second potential minimum $V''(z_1)=0 $, and the point at which both wells are of equal depth, $V(z_1) = V(0)$. The green line corresponds to the  second order phase transition at $z=0$. It is `hidden'  in the sense that any adiabatic annealing schedule progressing from right ($\Gamma =1$) to left sides will initially encounter the $1^{\text{st}}$ order transition from the $z=0$ minimum to $z_1>0$; the green curve never crosses the blue-orange bounded region, for all $\kappa$.}
\end{figure}
The penultimate step is to identify a shift operator:
\begin{equation}
\psi(z \pm \hbar) = e^{\pm \hbar D} \psi(z) =  \exp \left \{ \pm \hbar \frac{d}{dz} \right \} \psi(z) 
\end{equation}
in terms of the differential operator $D = d/dz$, the generator of translations in one dimension. 

\begin{figure}
	\includegraphics[width=3.0in]{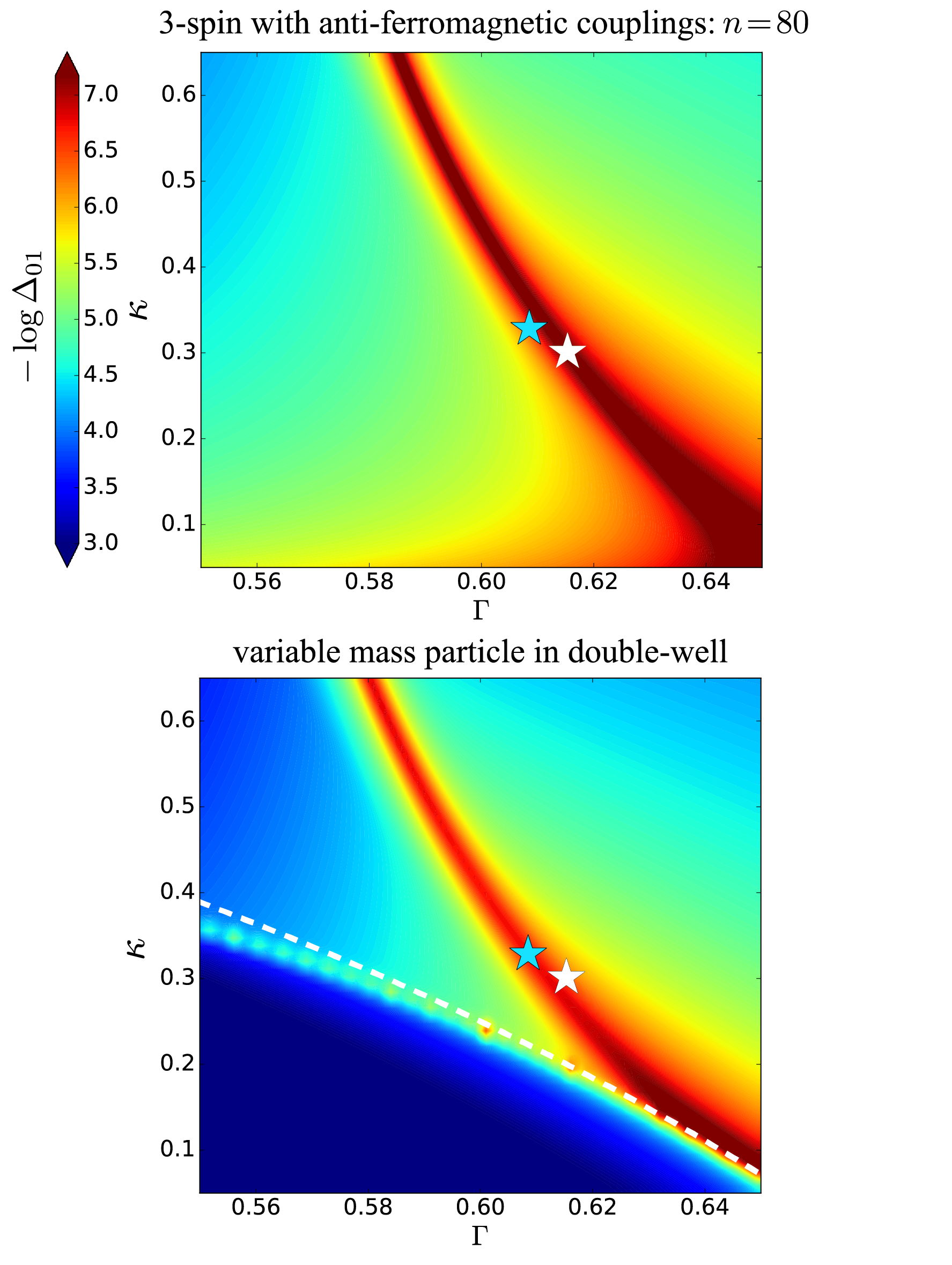}
	\caption{\label{compare_model}
		For direct comparison, here are plotted the spectral gap `landscape' for both systems, as a function of the control parameters $\Gamma$  (transverse field)  and $\kappa$ (anti-ferromagnetic driver). The upper plot depicts, for $n=80$ spins, the $3$-spin ensemble. The lower figure is the equivalent analytical model: a particle of variable mass moving in a one-dimensional continuous potential well. This 1D model is seen to exhibit the same qualitative features as the original spin ensemble (upper plot), except for an infinite mass singularity at $z=0$ coincident with the hidden second order phase transition (white dashed curve in lower plot), see eqn.\eqref{var_mass}. The saddle point of the phase transition ridge for the original $80$ spin system is indicated (in both plots) by a white star marker, and with  a cyan marker for the continuous model. Ground state wavefunctions associated with both models are shown in  FIG. \ref{optpot}.}
\end{figure}

Finally,  we expand everything to second order in the small parameter $\hbar$. This is only valid when the quantum state and the form of the potential are sufficiently smooth: $ \hbar^2 D^2 \ll \hbar D   \ll 1$, which may not always be the case. There are errors associated with truncation to $O(\hbar^2)$, but these should be less significant for larger qubit ensembles $n = 2 j  \gg 1$, resulting in the effective $\hbar \ll 1$. Then combinations like $\psi(z +\hbar) + \psi(z-\hbar)$ map to $ \cosh (\hbar D ) \psi(z) \approx [ 1 + (\hbar D )^2/2 ] \psi(z) $. As a consequence, transverse field term $\hat{J}_z $, as well as contributing to the  potential energy, is the origin of a kinetic energy term $- \frac{\hbar^2}{2} \frac{d^2 \psi}{dz^2}$ in the Schr\"odinger equation. In some sense, the transverse field \emph{provides} the kinetic energy that allows the quantum system to migrate through barriers.

All transverse terms of form $\hat{J}_x^q$ contribute to both kinetic and potential energy terms in the Hamiltonian:
\begin{equation}
\langle m | \left( \frac{\hat{J}_x}{j} \right)^q | \Psi \rangle \mapsto  \left[  \frac{q \hbar^2}{2} D(1-z^2)^{\frac{q}{2}} D  + (1-z^2)^{\frac{q}{2}} \right] \psi(z)
\end{equation}
Note the slightly unusual form of the Kinetic Energy operator for a variable mass, written in a manifestly Hermitian form:  $\hat{P} \hat{M}^{-1} \hat{P}/2$ (although such a position-dependent mass does occur in the semiconductor tunnelling literature \cite{mains1989effect}). Here, momentum operator $\hat{P} = -i \hbar D$ and inverse mass 
\begin{equation}\label{inverse_mass}
M^{-1}(z) = - q (1-z^2)^{\frac{q}{2}}.
\end{equation}
The potential energy contribution to $V(z)$ from the above mapping of $\hat{J}_x^q$ is $+(1-z^2)^{\frac{q}{2}}$.  An analytical treatment of the Schr\"odinger equation with position-dependent mass was presented in  Ref. \onlinecite{gonul2002exact}.

Now we are at a point we can write out the eigen-equation $\hat{H} | \Psi_k \rangle = E_k  | \Psi_k \rangle$ reformulated for a single particle of variable mass in a continuous potential: 

\begin{equation}\label{schrod1d}
\left[ \frac{1}{2} \hat{P} \hat{M}^{-1} \hat{P} + V(z) \right] \psi_k (z) = \frac{E_k}{\Gamma} \psi_k(z) \: ,
\end{equation}
The inverse mass operator can be zero or negative in the parameter space of $\{ \Gamma, \kappa\}$:
\begin{equation} \label{var_mass}
\hat{M}^{-1} (z)  = \sqrt{1-z^2} - 2 (1-z^2) \left[ \frac{1 - \kappa}{\gamma} \right] 
\end{equation}

Despite the fact this was a completely different approach to the mean field/classical spin derivation, the potential energy $V(z)$ coincides with the free energy in the mean field picture, eqn.\eqref{meanfield_VZ}. The key improvement is that in addition to defining a potential energy surface, we now have an analytical expression for the kinetic energy. Variable mass problems are interesting in their own right, and studying this one, with its possibility of infinite and negative mass, may reveal new behaviours within the p-spin paradigm. Or these anomalies may point to limitations of a model that is only quadratic in momentum $\hat{P}$. (Interestingly, the negative mass boundary is coincident with the second order phase transition in the p=3 case.) It is also intriguing how this model will behave for mesoscopic values of $n$: sufficiently large to maintain the validity of the transformation to continuous variables, but small enough that the effective $\hbar = 1/j$ is of a size that the ensemble behaves in an \emph{extravagantly} quantum manner. This is the opposite extreme to the thermodynamic limit $n \sim  \infty$, i.e. the classical limit $\hbar \sim 0$  where quantum effects vanish. The energy surfaces $V(z)$ for $z \in [-1,1]$ are indicated for a range of annealing parameter $\Gamma$, two snapshots taken at values $1$ and $0.4$ for $\kappa$, in FIG.\ref{PT} (a) and (b). The third subplot (c) will be discussed in section \ref{LMG}.

\begin{figure}
	\includegraphics[width=3.1in]{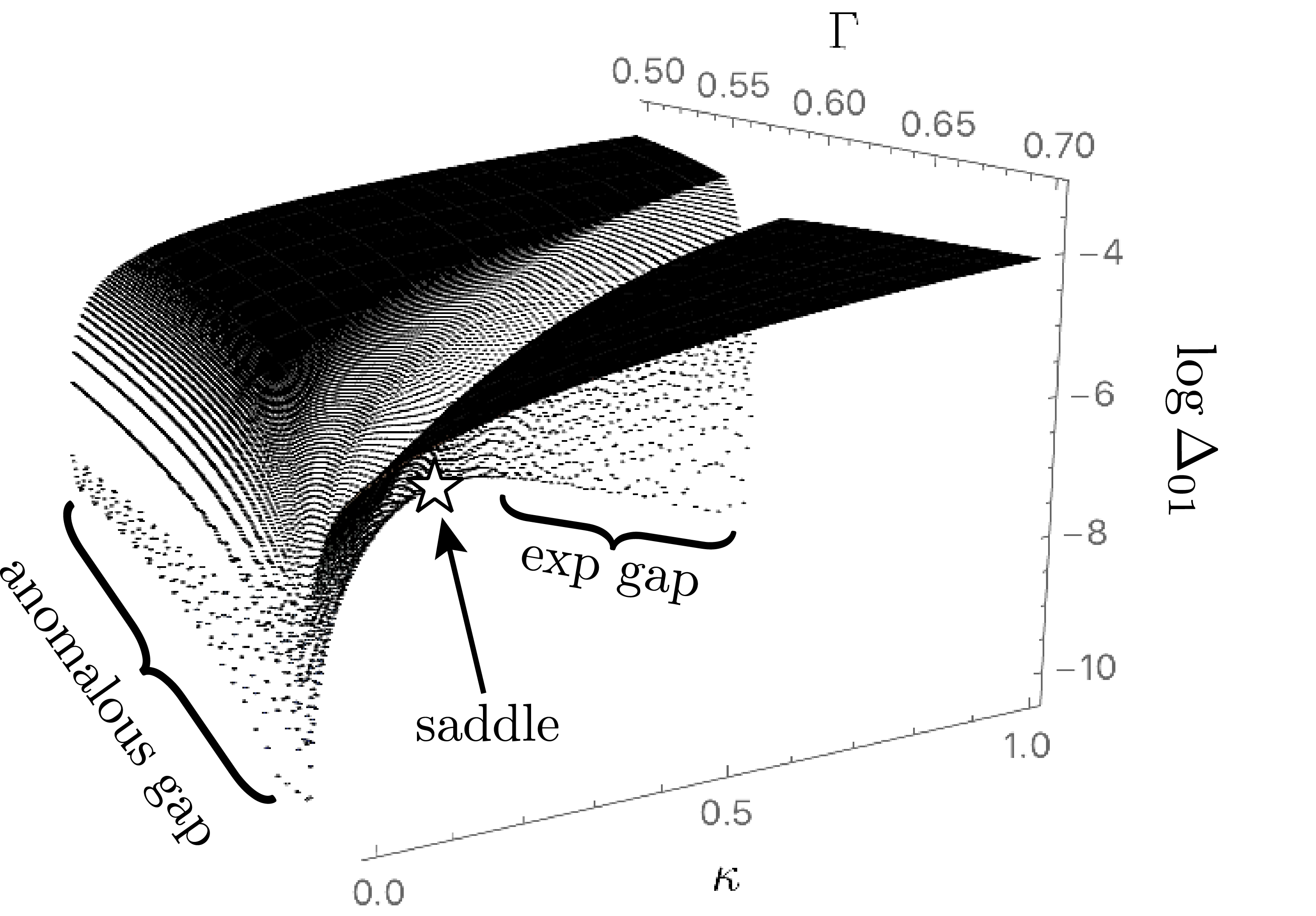}
	\caption{\label{maxmingap}
		The phase transition ridge features a saddle point, or `maximum minimum gap', more clearly visible in 3 dimensions when plotted against control variables $\kappa$ and $\Gamma$. The energy gap of this saddle point scales polynomially with system size $j$. At larger $\kappa$ (reducing the influence of the non-stoquastic driver) the gap size shrinks to be exponentially small in  $j$. For very small $\kappa$ the gap again quickly closes from polynomial to exponential or factorial, this anomalous gap was first discussed in Ref.\onlinecite{tsuda2013energy}.}
\end{figure}

\section{Adiabatic Evolution Across the Softened Phase Transition}

For adiabatic evolution the ground state will remain in the global minimum of the potential surface. The system begins at $\Gamma= 1$ in the unique minimum at $z=0$, then a second minimum-maximum pair are born as the transverse field is slowly turned off. The annealing ratio  at which the new minimum appears is given in eqn.\eqref{gamma_zero}. The coordinate location of the new minimum is rather clumsy if expressed as $z_1(\Gamma,\kappa)$, but it can be compactly expressed as a condition in terms of the associated polar angle $\theta_1$:
\begin{equation}
\sin (\theta_1 ) = \frac{\gamma  \sec (\theta_1 )+2 \kappa -2}{3 \kappa }
\end{equation}

This minimum $V(z_1)$ begins at a higher energy than the $z=0$ paramagnetic minimum but sinks quickly as $\Gamma \mapsto 0$, see FIG.\ref{PT} again. During the annealing this ferromagnetic minimum drops lower than the central minimum, and it is at this point the  ground state of the system jumps discontinuously to the ferromagnetic state in a first order phase transition. The point on the annealing schedule that corresponds to the minimum spectral gap actually occurs somewhere between the birth of the second ferromagnetic minimum and the point at which the wells have equal depth. In fact all the interesting quantum behaviour of this model occurs between these extremes, outside of which a mean field description will suffice. This is illustrated in FIG.\ref{gapstar},  a contour plot of the $j=40$ inverse gap $1/\Delta$ that indicates clearly a phase transition region bounded by curves associated with the birth of the second minimum  and the classical first order phase transition, in orange and blue respectively.  FIG. \ref{compare_model} compares this annealing landscape of the original $n$-spin ensemble and the corresponding 1D particle model we have developed, with good agreement.  Finally, FIG. \ref{polarplot2} in the appendix illustrates the domain of applicability of the mean field model quite explicitly (including its failure in proximity to the phase transition).

As a side remark, we mention that in the $\{\Gamma, \kappa \}$ parameter space, for  very small $\kappa \sim 0$ the gap closes very fast  to be factorially or exponentially small, without any associated phase transition -- this anomalous behaviour has been documented previously in Ref. \onlinecite{tsuda2013energy} and relates to the fact a finite number of spins $n = 2 j$ cannot exactly represent an irrational value of the variable $z = m/j$.  (In the current context this is not an interesting limit because the problem Hamiltonian vanishes when $\kappa = 0$.) See FIG.\ref{maxmingap} for a three dimensional visualization of the spectral gap landscape.

\begin{figure}
	\includegraphics[width=2.6in]{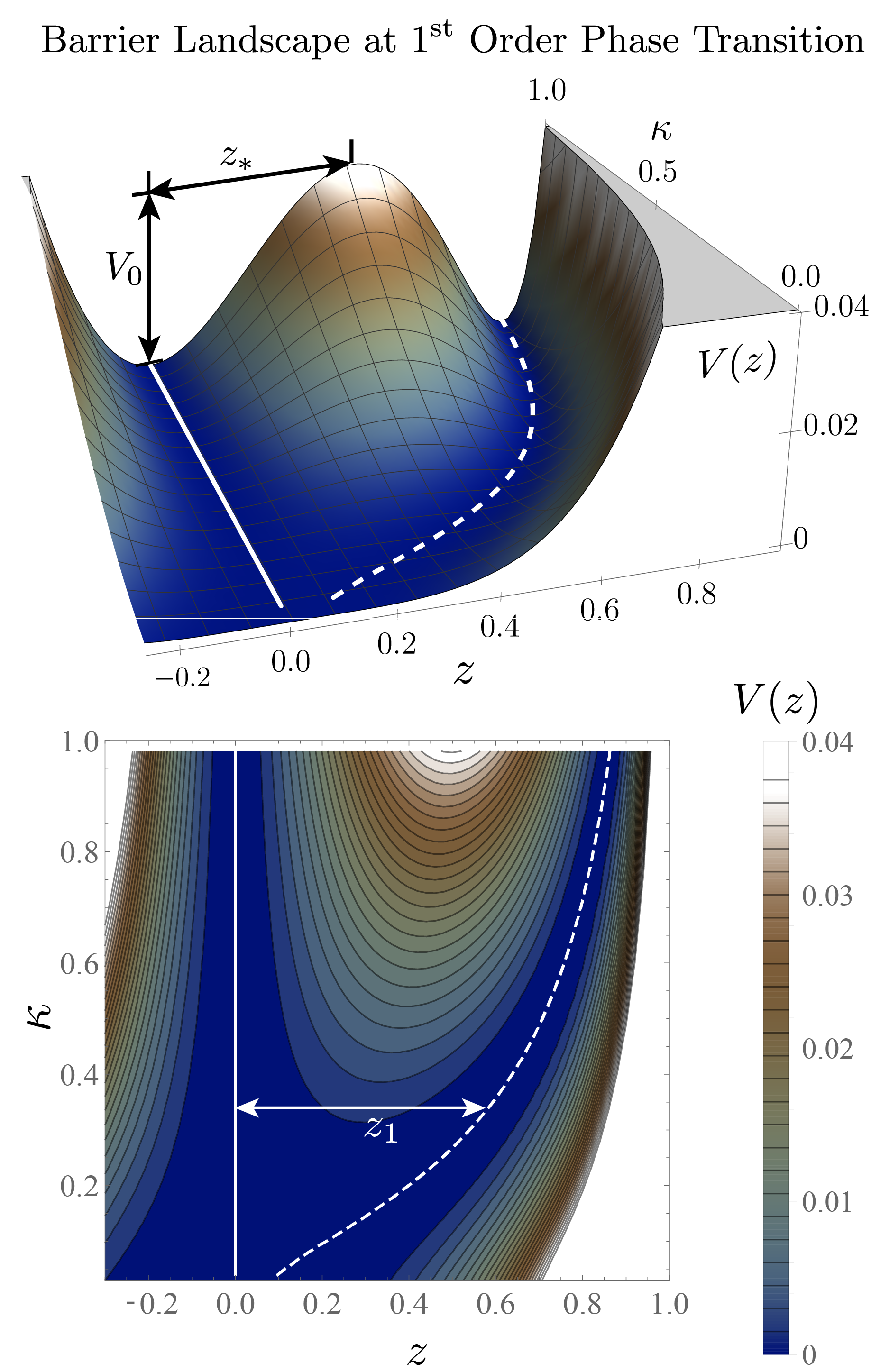}
	\caption{\label{barrierk}
		For the quantum $3$-spin model, the Hamiltonian contribution of the transverse anti-ferromagnetic  driver $+ (1-\kappa) \hat{J}_x^2$  is illustrated near the classical first order phase transition (wells are of equal depth). Control parameters $\kappa$  (indicated) and $\Gamma$ (not shown) and are chosen to fulfill this condition, and as $\kappa \mapsto 0$ the contribution of the non-stoquastic driver increases, lowering the barrier $V_0$, and reducing the separation of the potential wells $z_1$. The paramagnetic well is centered on the origin  (unbroken white line). The dashed white line describes the ferromagnetic minimum at $z=z_1$. Asymmetry of the wells is apparent even when they are of equal depth, as above.}
\end{figure}

In FIG. \ref{gapstar} along the ridge of the phase transition there is a white star marker indicating the location of the saddle, or \emph{maximum} minimum gap. As $j$ increases we would like to know whether this optimized gap begins to shrink at a polynomial or exponential rate. In terms of classical phase transition analysis, the second order phase transition (green arc of FIG. \ref{gapstar} lower plot) is always hidden behind the first order transition (blue curve), and therefore there is no hope to produce a polynomial-sized gap. Stated using carefully chosen language in Ref. \onlinecite{nishimori2016exponential}  -- on examining the potential surface, \begin{quotation} \emph{`the first-order transition line persists down to $\kappa =0$ . This fact may be interpreted in terms of the Landau theory of phase transitions that there would appear a cubic term in the Landau free energy for the cubic Hamiltonian with p = 3, which strongly enhances the possibility of first-order transition.'}\end{quotation}.

Here the authors are referring to the cubic term that arises in eqn.\eqref{meanfield_VZ}, setting $p=3$. The apparent inevitability of barrier penetration via tunnelling, with the associated exponentially small spectral gaps and first order phase transitions in many quantum annealing landscapes (beyond simple p-spin models, e.g. spin glasses) is a phenomenon much-cited in arguments against the efficiency of the adiabatic algorithm for practical problems \cite{jorg2010energy}. This is somewhat ironic: tunneling itself is one of the vaunted traits of quantum annealing that offer it an advantage over classical algorithms. Let's examine what happens to the double well as $\kappa$ is reduced, turning on the the influence of the non-stoquastic antiferromagnetic driver $+ \hat{J}_x^2$.

From FIG.\ref{barrierk} it's seen that reducing $\kappa$ also lowers the barrier, it becomes completely suppressed only for $\kappa = 0$. This limit, however, brings us back to the anomalous case discussed earlier\cite{tsuda2013energy}. How does the optimal non-stoquastic driver contribution $\kappa_c$ scale with the system size $n = 2 j$? If $\kappa_c$ approaches zero too quickly it suggests that the non-stoquastic terms may not be very useful for larger ensembles. More crucially, what gap scaling can we achieve, even without being able to suppress the barrier entirely?

Algebraic analysis of the $\kappa \ll 1$ limit (appendix \ref{appendix_kappa}) shows at small $\kappa$ (strong non-stoquastic driving) the coordinate distances $z_1$, characteristic frequencies $\omega$ and barrier heights $V_{0}$ scale  $ \mapsto \{\kappa, \kappa^2, \kappa^4\}$, respectively. And application of Rayleigh criterion from eqn.\eqref{Rayleigh} indicates a scaling law at the saddle point:
\begin{equation}\label{optkap}
\kappa_c  \sim O( \sqrt{\hbar}) =  \alpha / \sqrt{j} \; .
\end{equation}
where  $\alpha$ we shall call the `Rayleigh coefficient'.
\begin{figure}
	\includegraphics[width=2.8in]{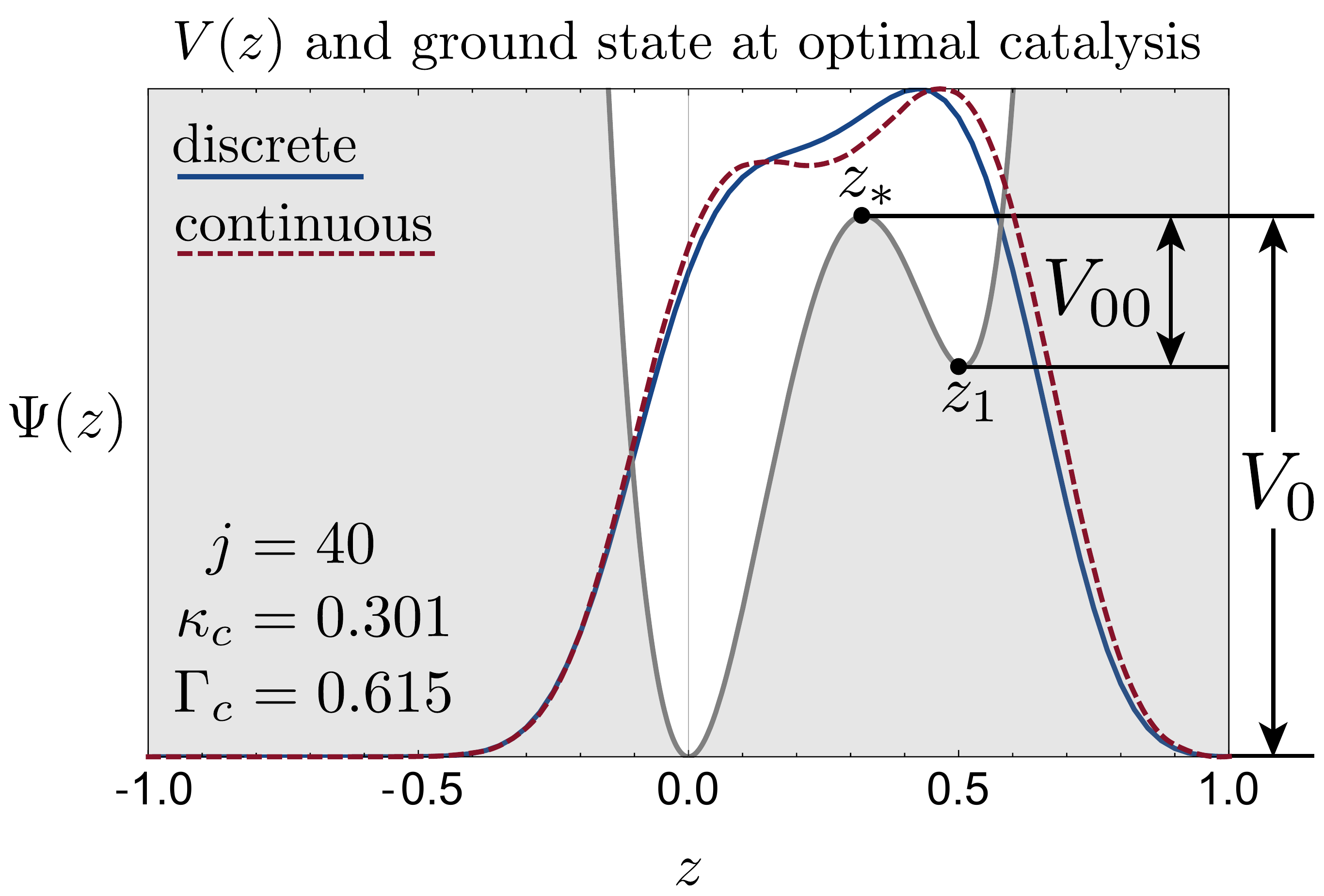}
	\caption{\label{optpot}
		The Rayleigh limit is clearly violated at optimal catalysis for both the original discrete spin system, and the continuous model  -- i.e. for the widest minimum gap $\Delta_{01}$ -- here illustrated for $n=80$ qubits. Optimal control parameters $\{\Gamma_c, \kappa_c\}$ were found for the spin system by numerical search using a truncated Newton method. The ground state $\Psi(z)$ (blue and dashed red curves) no longer has isolated components in each well, they are largely coalesced. Note the asymmetry, not only of the potential well $V(z)$ (grey filled curve), but of the wavefunction, biased towards the ferromagnetic well (centered on $z_1$). }
\end{figure}

This is encouraging, firstly because it vindicates our choice to work in the limit of small $\kappa$ for larger ensembles. Also, it says the relative strength $\kappa$ of the problem Hamiltonian $- \hat{J}_z^3$ to the non-stoquastic driver $+ \hat{J}_x^2$ must decay only polynomially in system size $n = 2j$. An optimal barrier will decay as $\kappa_c^4 \propto 1/j^2$. For any finite ensemble an optimized catalysis occurs for a non-zero barrier. Total barrier suppression, if it were even possible, would be sub-optimal. (It is of course, not possible in the $3$-spin model.)

Combining eqn.\eqref{hbar_n} with eqn.\eqref{optkap} above, $\hbar =1/j \propto \kappa_c^2$ for optimal catalysis,  the vacuum energy for the isolated wells $\hbar \omega_{0,1}/2$ is of same order as the potential barrier height $V_0 \sim \kappa_c^4$; all energy scales are equivalent.

Let's examine the Rayleigh limit $\xi_{1} \sim \beta$ for the asymmetric potential e.g. of  FIG.\ref{optpot}. We may apply the phase transition resonance condition  that was introduced in section \ref{resonance_section}. Mapping the non-stoquastically driven 3-spin into the piecewise-parabolic  potential produces:
\begin{equation} \label{rayleigh_answer}
\{\xi_{1}, \beta_1, \xi_2, \beta_2\} \mapsto \left\{ \frac{\alpha}{3^{1/4}}, \: 1,  \: \frac{\alpha}{3^{1/4}},  \: \frac{1}{3^{1/4}}\right\}
\end{equation} 
where $\alpha$ is the scaling coefficient of the Rayleigh criterion, to be recovered presently  ($\kappa_c = \alpha / \sqrt{j}$). Asymptotic expressions for $\xi, \beta$ terms in the small $\kappa$ limit are also worked out in appendix \ref{appendix_kappa} rather than interrupting the current narrative. The asymmetric potential in  scale-free coordinates  was depicted in FIG.\ref{antirayleigh}. 

\begin{figure}[b]
	\includegraphics[width=2.2in]{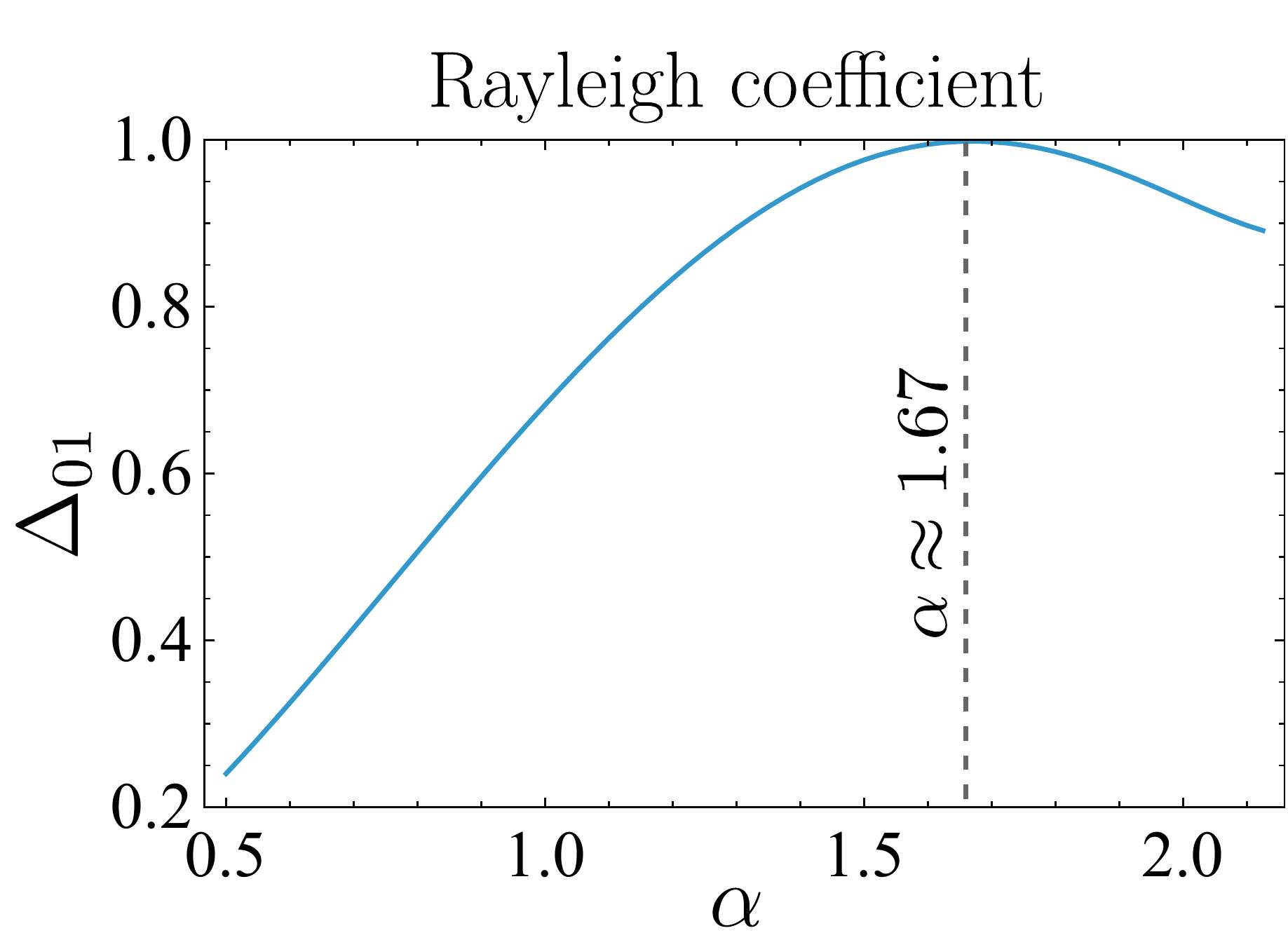}
	\caption{\label{ray}
		The spectral gap $\Delta_{01}$ in the 3-spin is formed by multiplying the $O(1)$ gap in scale-free coordinates by $\hbar \omega_{*}$, which produces $\Delta_{01} \sim  O(1/j^2)$. To find the numerical coefficient we may apply the delocalization resonance condition of section IV  to produce this curve of the gap (in arbitrary units) as a function of the Rayleigh coefficient $\alpha$. It features a maximum at  $\alpha \approx 1.67$, corresponding to the saddle of the phase transition. }
\end{figure}

From eqn. \eqref{gapeqn2} the energy $\hbar \omega_{*}$ contributes a multiplicative factor $\alpha^2$ to the gap $\Delta_{01}$, using the small-$\kappa$ result of eqn. \eqref{freqs}. The saddle point of the phase transition (maximum minimum gap) occurs when the spectral gap times $\alpha^2$ is a maximum:
\begin{equation}\label{rayleigh_two}
 \kappa_{c}  \approx\frac{1.67}{\sqrt{j}} \:  ,
\end{equation}
the explicit dependence of the gap on $\alpha$ is plotted in FIG.\ref{ray}. (The numerical solution to the scale-free problem unlocks the fundamental scaling coefficient, universal to the 3-spin problem of any size $j \gg 1$.)

\begin{figure}
	\includegraphics[width=3.in]{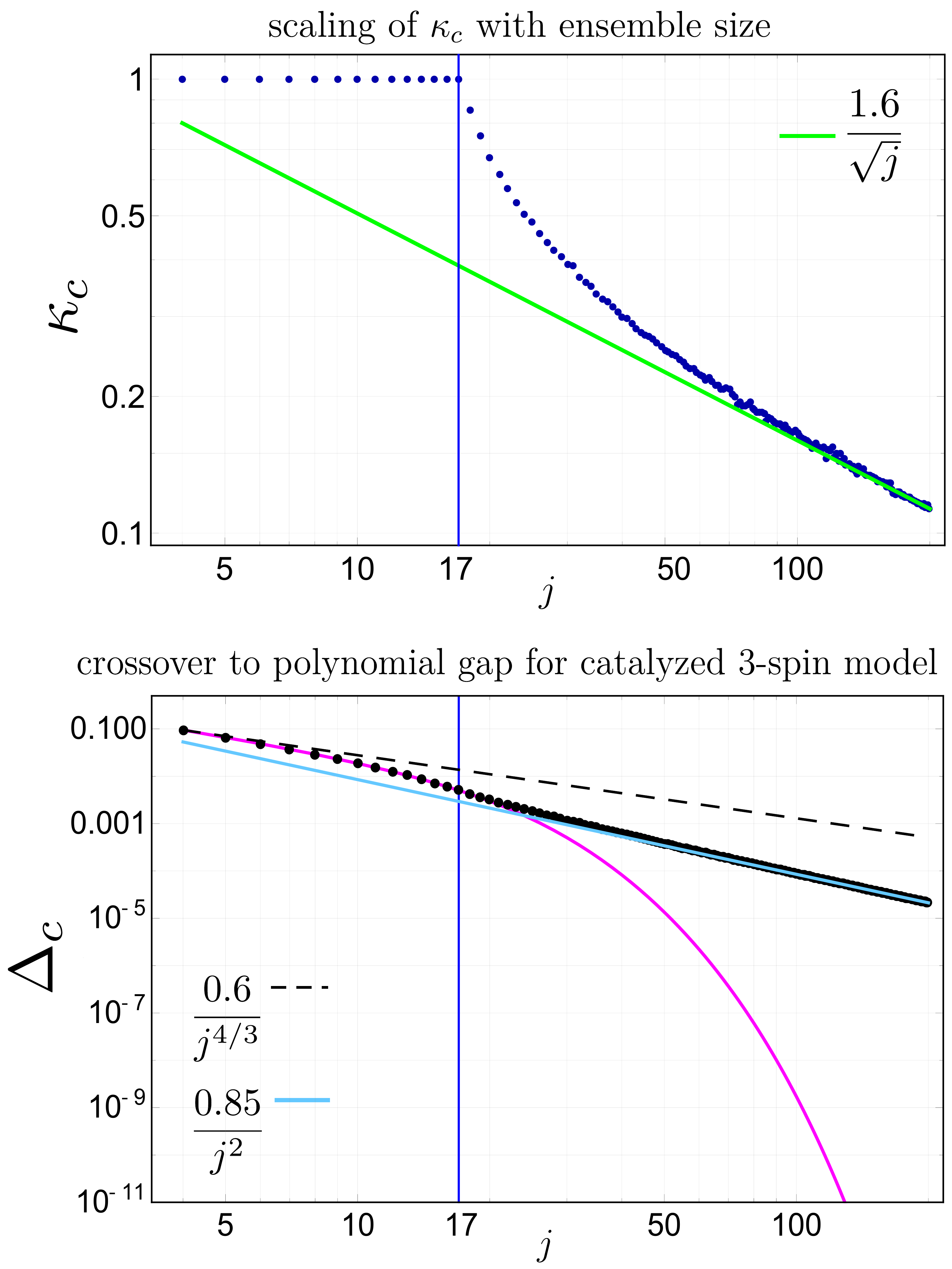}
	\caption{\label{kscaling}
		Upper plot collects numerical results for the non-stoquastic control parameter $\kappa_c$ of  $3$-spin system for $4\leq j \leq 196$. The optimized parameter values asymptote to $\kappa_c \propto 1/\sqrt{j}$, eqn.\eqref{rayleigh_two}. For systems of $j \leq 17$ apparently the ground state is always delocalized over the barrier, because the effective $\hbar = 1/j$ (and therefore vacuum energy) is  large. In these cases no catalysis (no barrier suppression) is necessary to produce a polynomial gap: $\kappa_c =1$. The lower plot shows the scaling of the minimum gap, also with system size. Even though it is forbidden from the perspective of classical phase transition theory, a polynomial gap can be maintained in the quantum $3$-spin model by suppressing the potential barrier to an optimal non-zero height $V_0 \propto \kappa_c^4$. A partially lowered barrier permits the two well components to merge;  at the Rayleigh limit they become a unimodal state, with polynomial spectral gap. Without any catalysis the minimum gap will quickly close at an exponential rate (magenta line), the crossover again indicated at  $j \approx 17$. The exponential scaling of the uncatalyzed first order transition asymptotes to $\Delta \sim \exp \{- 0.175 j \}$, as documented in Ref.\onlinecite{jorg2010energy}. The optimal gap (black dots) eventually converges on the  asymptote (cyan line) predicted by the continuous model, eqn.\eqref{gap_law}. The rate of gap closure is always faster than that of the Lipkin Meshkov Glick ($2$-spin) model\cite{Fallieros1959,lipkin1965validity}   $\Delta_c \propto j^{-4/3}$, a continuous phase transition. Its scaling is indicated in the lower plot by a dashed line. }
\end{figure}

Going back to, and comparing, the original spin system, numerical results for  $ n \lesssim 400$ are presented in the upper plot of FIG.\ref{kscaling}  which also asymptotes to  $\kappa_c \sim 1.6/ \sqrt{j}$, confirming the validity of the variable-mass model and the simplification to a piecewise-parabolic potential.

These results further suggest that smaller systems will more easily violate the Rayleigh separation criterion. In the thermodynamic/classical limit, $j \sim \infty$ it is impossible to approach this Rayleigh boundary. Classically, one will always have a first order phase transition and exponentially small gap. More correct than `classical',  we might designate this the `large spin limit';  the framework we have illustrated remains quantum mechanical and tunnelling is permissible, if unlikely.

Referencing eqn.\eqref{gapeqn2} we know that $\Delta_{c} \sim O(1/j^2)$, with one power of $1/j$ coming from $\hbar$ and the other from $\omega_{*}$, the latter was calculated at the critical $\kappa_c$ in eqn. \eqref{freqs} . The scale-free analysis that produced $\alpha \approx 1.67$ also provides the scaling here:
\begin{equation}
\Delta_c\sim \frac{ \sqrt{3}}{\: 2 j^{2}} \; . \label{gap_law}
\end{equation}
This compares remarkably well to the original 3-spin system, verified numerically to  $400$ qubits in the lower plot of FIG.\ref{kscaling}.

The prediction of a crossover from exponentially small gap to a polynomial one and the resulting quantum speed-up, especially in models where it was assumed not possible, is a central result of this note -- as is the presentation of an optimized catalysis  (energy barrier suppression) associated with a type of Rayleigh criterion and resonance for the quantum ground state of a double well. A peak in mobility is  possible because of the competition between the increasing mass and increased localization (narrowing of potential well) of the state that occurs at lower $\kappa$. The former decreases energy scales and the latter increases them. The inclusion of kinetic energy and quantum uncertainty in the analysis may require redrawing of the boundaries in many phase diagrams produced for models such as p-spin, that had previously been based on consideration of the classical potential surface alone.

\section{\boldmath$t^{*}$ : Time to Solution Via Optimal Path}

Earlier we gave a simple justification in the tunneling case that time to solution and minimum gap at the phase transition are inversely related: $t^{*} \sim 1/\Delta_{01}(\Gamma^{*},\kappa^{*})$. Now we have a Hamiltonian evolving under the influence of two drivers, $\Gamma$ and $\kappa$, for which we can adapt a recipe presented in Ref.\onlinecite{van2001powerful}. With gradient operator $ \nabla = \left( \frac{\partial }{\partial  \Gamma} ,\frac{\partial }{\partial  \kappa} \right)$:
\begin{equation} \label{pathfinding_eqn}
t^{*} = \frac{1}{j}  \int_{C}  \frac{1}{\Delta^{2}_{01}} \left|\left| \nabla \hat{H} \right|\right|_{2} . \: (d \vec{C})
\end{equation}
% \left|\left| \frac{\partial \hat{H}}{\partial  \Gamma}\right|\right|_{2} (\text{d} \Gamma) + \left|\left| \frac{\partial \hat{H}}{\partial  \kappa}\right|\right|_{2} (\text{d}  \kappa)
The open curve $C$ connects initial point $\{\Gamma, \kappa \} = \{1,1\}$ to final point $\{0,1\}$ in control space and $||M||_2$ denotes the $2$-norm of a matrix $M$. Approximately, for the Hamiltonian of eqn.\eqref{3spinwithkappa} in the limit $j \gg 1$ we have $||\partial  \hat{H} / \partial  \Gamma||_2 \sim 2- \kappa$, and $||\partial  \hat{H} / \partial \kappa||_2 \sim 1- \Gamma$.

We believe annealing through the saddle point, identified by a star in FIG.\ref{gapstar} permits a fast (polynomial time) adiabatic evolution. Until now we have seen only the $1/j^2$ scaling of the gap size at the saddle; let's establish that optimal paths actually traverse the phase transition in the neighborhood of this saddle.

\begin{figure}
	\includegraphics[width=2.9in]{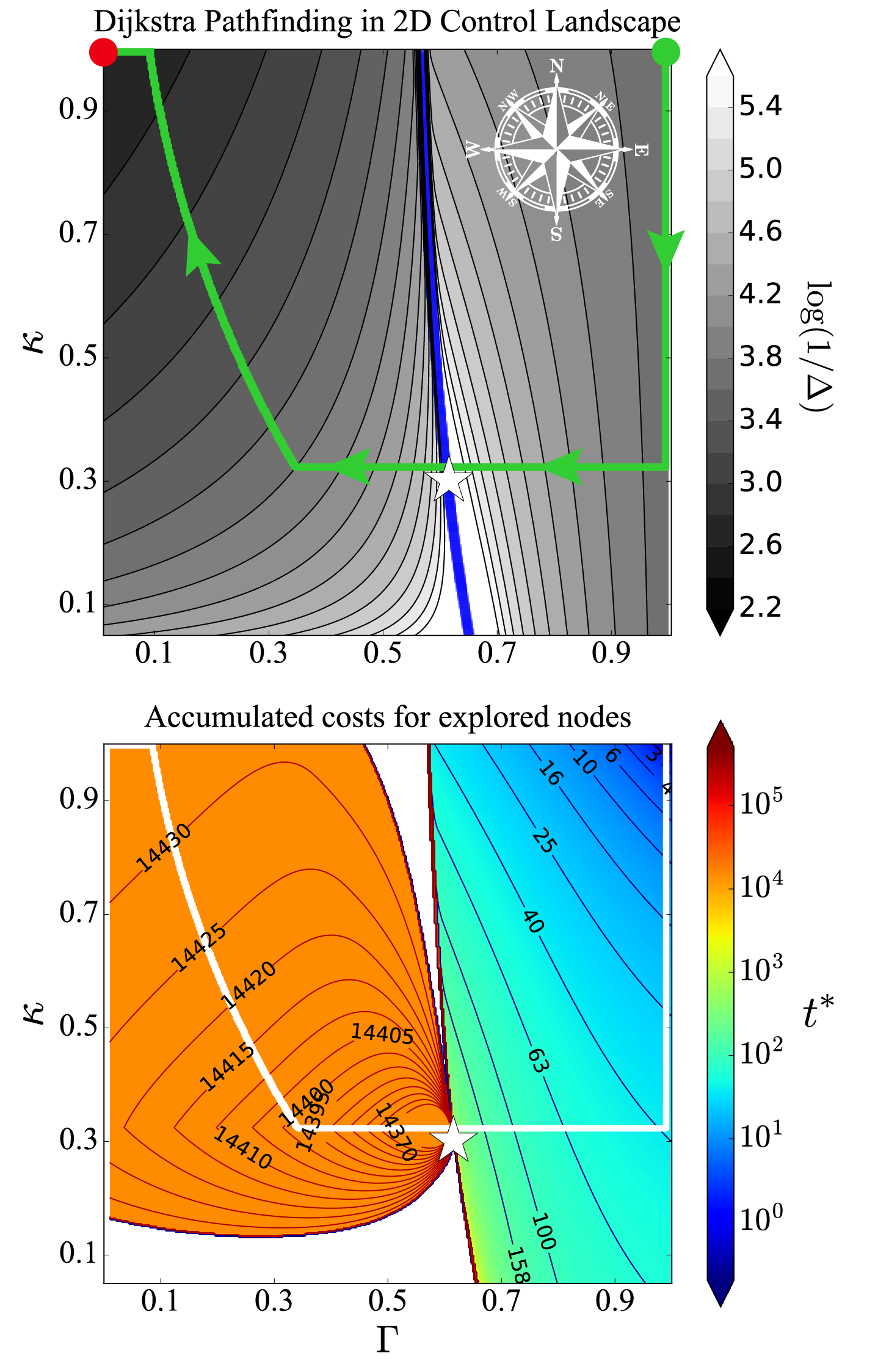}
	\caption{\label{dijkstra_fig}
		The algorithm due to Dijkstra is employed on a graph produced by rasterizing the control landscape of FIG.\ref{gapstar}, here the ensemble size $j=40$. Contours here are also of $\log 1/ \Delta_{01}$, the cost of moving between adjacent nodes  $\propto 1/ \Delta^2_{01}$. Darker-shaded regions may therefore be more rapidly traversed. Movement costs along graph edges, between pixels, are defined by eqn. \eqref{pathfinding_eqn}. The green line represents an optimal adiabatic contour $C^{*}(\Gamma,\kappa)$ from the initialization parameters $\{ \Gamma, \kappa \} = \{  1, 1 \}$ (green dot) to the final values $\{  0, 1 \}$ (red dot). The blue pixels of the upper plot represent the phase transition ridge, with the saddle  indicated by a white star marker. One might imagine an explorer journeying south from his home in the north-east corner to cross a river close to the shallowest point before heading to his destination in the north-west. There are different anisotropic frictional movement costs for going south $(1- \Gamma)(\delta \kappa)$ versus west, $(2- \kappa) ( \delta \Gamma)$, independent of the node cost. This explains the zero friction path  traversed due south from $\{1,1\}$ to $\sim \{1,0.3\}$ that then turns westward towards the saddle marker. On the lower plot  is  $t^{*}$, the cumulative contribution to the total adiabatic time. Unexplored regions are left uncolored in the lower plot. Admittedly, there will be error in the sampling of the phase transition terrain (the ridge is basically a delta function) which might be improved by adaptive sampling of regions with steep gradients.}
\end{figure}

To investigate such paths $C$ we can rasterize the contour landscape of FIG.\ref{gapstar}, turning it into a grid of pixels. Each pixel becomes a node on a graph, we can use eqn.\ref{pathfinding_eqn} to understand movement costs (time penalty) along edges connecting these nodes. Restricting movement to the $\{N, S, E, W \}$ directions (diagonal costs are not uniquely defined) mean that each node/pixel is connected to at most four others. Next, we can employ a pathfinding algorithm such as that pioneered by Dijkstra in the late 1950s, Ref.\onlinecite{dijkstra1959note}, which uses a prioritized queue to explore the graph. The algorithm is greedy, partial paths are favored that have the lowest accumulated costs. Shortest paths found in this manner are presented in FIG. \ref{dijkstra_fig}. We observed in FIG.\ref{kscaling} that the saddle moves off the straight-line path ($\kappa = 1$) connecting $\{0, 1 \} \leftrightarrow \{1,1 \}$ for $j >17$. In contrast, the pathfinding algorithm finds an optimal route that deviates from the beeline trajectory for $j \gtrsim 20$. The optimal path ventures close to the saddle point (akin to a mountain pass through the  phase transition ridge) only for larger $j$ -- for instance the case $j=40$ presented in FIG.\ref{dijkstra_fig}.

\begin{figure}
	\includegraphics[width=3.in]{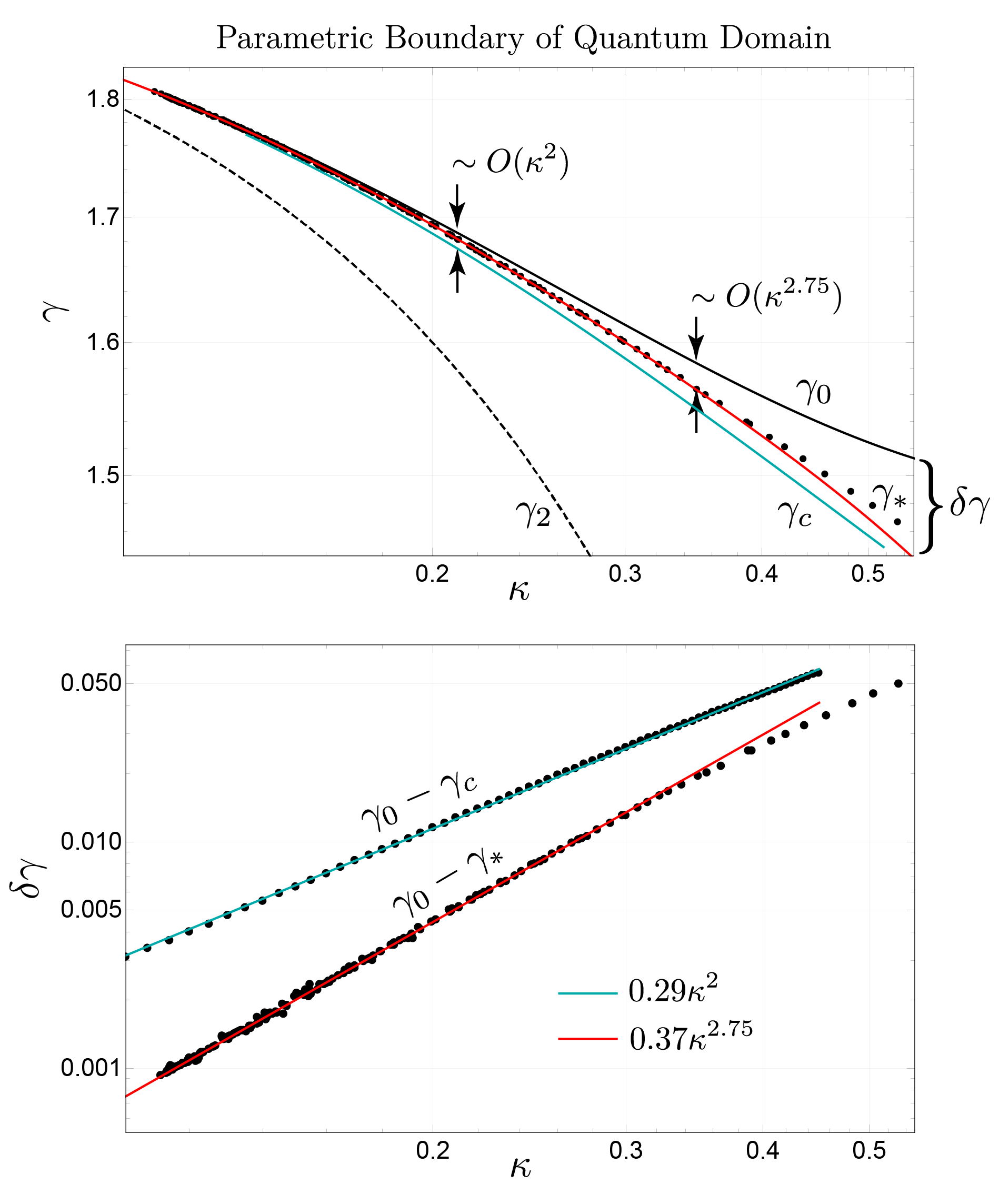}
	\caption{\label{gamma_scaling}
		In $\{\kappa,\gamma\}$ parameter space, the quantum domain  exists between the birth of the ferromagnetic well in the potential at $\gamma_{0}$ and the point of the \emph{classical} first-order phase transition $\gamma_c$, where the potential wells have equal depth. Also shown is the hidden second order transition $\gamma_2 = 2(1- \kappa)$, where the paramagnetic minimum becomes a maximum (dashed line). Between $\gamma_c$ and $\gamma_0$ appears the true quantum phase transition at $\gamma_{*}$. Scaling of  $\gamma_{0} - \gamma_{*}$ is observed between $\kappa^2$ and $\kappa^3$; a fit line for $\kappa^{2.75}$ is indicated on a logarithmic scale in the lower plot  (red line).}
\end{figure}

To find an analytical answer to the scaling of this algorithm with $j$, we will make some approximations. First, let's assume the dominant contribution to $t^{*}$ will come from the vicinity of the phase transition. In essence we want to find an effective $\delta \Gamma$, or $\delta \gamma$ corresponding to the phase transition region. We can begin by assuming the optimal path segment $\delta \vec{C}$ will traverse the transition in a direction normal to the curve $\gamma_0$ from eqn.\eqref{gamma_zero} that defines the boundary of the quantum region. In the parameter space of $(\kappa,\gamma)$ the vector $\vec{\gamma}_0 = (\kappa, 2(1 - \kappa) + 9 \kappa^2/4)$ has a normal vector  $\vec{n} \approx (9 \kappa/2 -2 , -1)$. This produces
\begin{equation}
t^{*} \sim \frac{\delta \kappa}{j \Delta_{c}^2} \left\{ O(1) \right\}
\end{equation}
where we can easily find the effective width of the phase transition $\delta \kappa$ from the intersection of the normal line with curves $\gamma_{*}$ and $\gamma_0$, see FIG.\ref{gamma_scaling}.

The hidden second order phase transition (occurring when the paramagnetic minimum at the origin $z=0$ becomes a maximum) occurs precisely at $\gamma_{2} = 2 (1- \kappa) $, which is the same as $\gamma_0$ to to first order in $\kappa$. The quantum phase transition must occur between $\gamma_0$ and $\gamma_2$, and therefore  $t^{*} \sim O(\kappa^2)$ or smaller.

Numerical results of FIG.\ref{gamma_scaling} suggest that $\delta \gamma \sim O(\kappa^{2.75})$ at the critical $\kappa_{c}$. The associated small change $\delta \kappa$ normal to the phase transition ridge must also be:
\begin{equation}
\delta \kappa_c \sim O(\kappa^{2.75}) .
\end{equation}
when $\kappa \ll 1$. Putting all our scaling relationships together, including the Rayleigh limit $\kappa_c  \sim O(j^{-1/2})$ and gap size scaling $\Delta_c \sim O(j^{-2})$ gives
\begin{equation}
t^{*} \sim O(j^{\alpha})
\end{equation}
for the catalysed time complexity where we have given the analytical bound $\alpha < 2$ and numerical evidence for $n \in [1, 400]$ qubits indicates an $\alpha \sim 13/8$. 

\textbf{The overall algorithmic complexity of the catalysed 3-spin is polynomial, between linear and quadratic in the number of spins or qubits.}

The continuous model's validity relies on a degree of smoothness in e.g. the wavefunction and its derivative. We may be precluded from any refinement on smaller scales than $\hbar = 1/j$, e.g. in the coordinate $z$. Optimal catalysis has $\kappa_c \sim 1 / \sqrt{j}$ so sharp effects such as locating the phase transition ridge might be associated with higher powers than $\kappa^2$, and thus may not be captured here. For these reasons we may be restricted to the statement that in parameter space, the phase transition ridge has width $1/j$ \emph{or smaller} at the saddle point of optimal catalysis.

\section{Conclusions and Outlook}

One might expect that quantum mobility in a potential with tall barriers increases monotonically as those barriers are suppressed, by application of external control fields or couplings. This can result in an exponential speed-up in time to solution. Some coupled spin systems, however, exhibit a sweet spot, an optimal catalysis, where mobility depends on more than just complete barrier suppression. Indeed, that full suppression may not even be possible, and the exponential speed-up seems out of reach. For instance, the 3-spin model with non-stoquastic driver has a strong dependence on the variable mass that results in a saddle point of the barrier in parameter space. 

\begin{figure}[h]
	\includegraphics[width=3.3in]{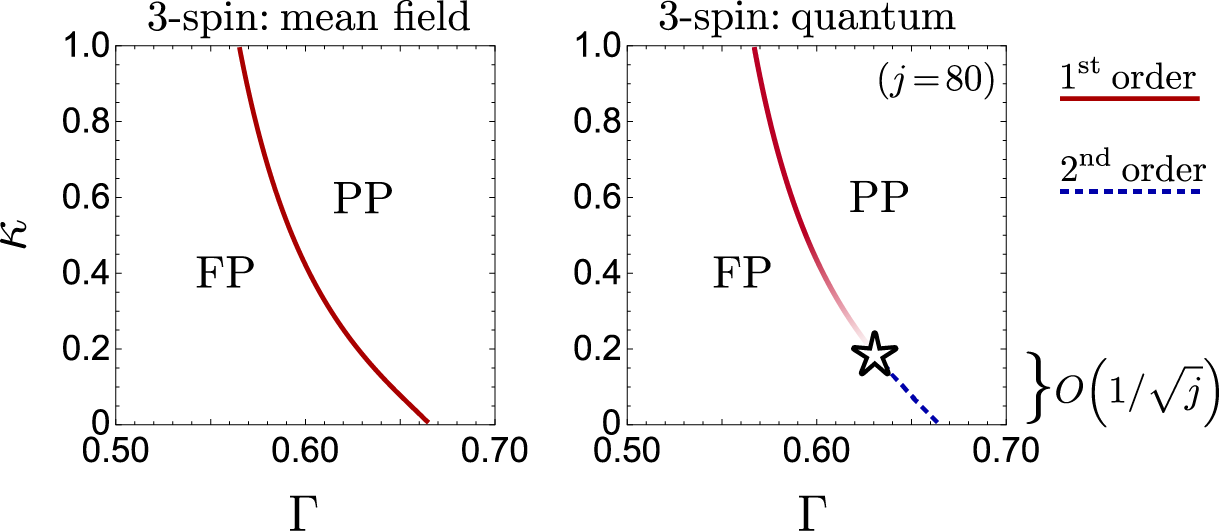}
	\caption{\label{portrait}
		A light at the end of the tunnel: Subject to a full quantum analysis, the phase portrait for the anti-ferromagnetically driven $3$-spin must be redrawn. The boundary between paramagnetic phase (PP) and ferromagnetic phase (FP) in the space of control parameters $\Gamma$ and $\kappa$ is no longer first-order, and admits a softer transition. This 'polynomial back door' of order $1/\sqrt{j}$ bounded by the saddle point (star marker) circumvents the exponential slowdown in time-to-solution associated with tunnelling; the annealling proceeds in polynomial time.}
\end{figure}

 We have showed that kinetic energy scales should not be ignored, as they are in mean field models. The scale is set by the effective Planck constant $\hbar = 1/j$, the reciprocal ensemble size. This `quantum uncertainty' dictates the ability of a ground state to delocalize across barriers. We re-purpose the Rayleigh optical separation criterion for quantum computing, and identify its violation as a harbinger of \emph{exponentially} enhanced mobility, see FIG.\ref{portrait}. 
 Also, when a double-well potential exhibits asymmetry we identify a resonance condition, allowing  the quantum phase transition point to be precisely located, as distinct from the nearby classical one. The full quantum treatment of annealing through shallow barriers can lead to radically different conclusions about time complexity of the algorithm.

To illustrate this `optimized catalysis' we created an analytical model, and verified numerically via a pathfinding algorithm, previously unexpected polynomial scaling (and universal coupling coefficients) of the time-to-solution for the quantum 3-spin catalyzed by an anti-ferromagnetic coupling. 

In terms of future work, if typical barrier heights are known in an annealing problem, e.g. $V_0 \sim n^{1/3}$ for some spin glasses, one might match vacuum energies to that scale in our models, to effect polynomial time solutions of otherwise `hard' problems. The challenge then will be the optimal control of the Hamiltonian landscape, without leveraging prior knowledge of minimum gap, saddle or barrier locations and magnitudes. Because the vacuum delocalization effect we describe relies on mesoscopic-scale systems, (a large effective $\hbar = 1/j$ balanced against larger problem instances $n = 2 j$) there is motivation to distribute large computations in an optimal way among smaller quantum sub-systems, of e.g. $10$ to $1000$ qubits.

These results hold in the adiabatic limit at zero temperature. For the non-zero temperature case, we should assume that $kT \ll V_0$; for if $kT \sim V_0,$ simulated annealing is known to be an efficient approach.

I would like to express my gratitude to Waleed Kadous at Uber, Zhang Jiang at Google, Itay Hen of the  University of Southern California, Birgitta Whaley, William Huggins and Norm Tubman, all at University of California, Berkeley, and finally, Sergey Knysh, my longtime collaborator at the NASA Ames Quantum Laboratory. My thanks to each for their encouragement, diverse perspectives and stimulating discussions as this work progressed. 

\appendix

\section{Details of the Piecewise-Parabolic Potential}\label{appendix_potential}

In order to understand vacuum delocalization we may write a Schr\"odinger equation for states in the vicinity of the barrier maximum, where we will assume the potential is dominantly quadratic. 
\begin{equation}
- \frac{\hbar^2}{2m} \frac{d^2 \psi^{+}_0}{d z^2} + \left[ V_0 - E^{+}_0  -  \frac{m \omega_{*}^2}{2} z^2\right] \psi^{+}_0 = 0 .
\end{equation}
It's possible that in this regime a typical WKB approach will fail -- although the WKB series is an exact asymptotic expansion, its truncation to leading  terms may not be justified here. (It should be noted that a successful WKB analysis on the phenomenon of quantum transport across a fully-suppressed barrier was presented in Ref. \onlinecite{bulatov2002total}.)

General characteristics become clearer by switching to non-dimensionalized energy and length scales:  
\begin{subequations} \label{rescaled}
	\begin{align}
	\xi &= z /\sigma_{*} \\
	\delta^{\pm} &= (V_0- E^{\pm})/(\hbar \omega_{*}) \\
	\epsilon &= V_0/(\hbar \omega_{*})  - \delta^{+}\\
	\beta &= \sigma_1/\sigma_{*} \\
	\xi_1 & = z_1/\sigma_{*}
	\end{align}
\end{subequations}

The rescaled energies $\delta^{\pm}$ are the energy deficit under the barrier summit for the ground and excited states respectively, and measured in `quanta' $\hbar \omega_{*}$ of the inverted maximum\cite{ford1959quantum}. The rescaled $\epsilon$ is the ground state energy measured from the well-bottom at $V = 0$. The barrier height will increase with the square of well separation $\xi_1$:
\begin{equation}
V_0 = \frac{\hbar \omega_{*}}{2} \left(\frac{1}{1+ \beta^4} \right) \xi_1^2 = \frac{\hbar \omega_1}{2} \left(\frac{\beta^2}{1+ \beta^4}  \right) \xi_1^2 .
\end{equation}
For fixed well separation $\xi_1$ the barrier is lowered monotonically for increasing $\beta = \sigma_1 /\sigma_{*}$, i.e. widening the wells also suppresses the barrier. We may rewrite the Schr\"odinger equation for the ground and first excited states in the vicinity of the barrier maximum (that is $ \xi$ and  $|\delta^{\pm}| \lesssim 1$) as:
\begin{equation}
\label{gap-pair-eqns}
- \frac{d^2 \phi^{\pm}_0}{d \xi^2} + (2 \delta^{\pm} -\xi^2) \phi^{\pm}_0  = 0 
\end{equation}
enabling us to `roll up' parameters like $m, \hbar, \omega$ into the new variable $\xi$.

\begin{figure}
	\includegraphics[width=3.2in]{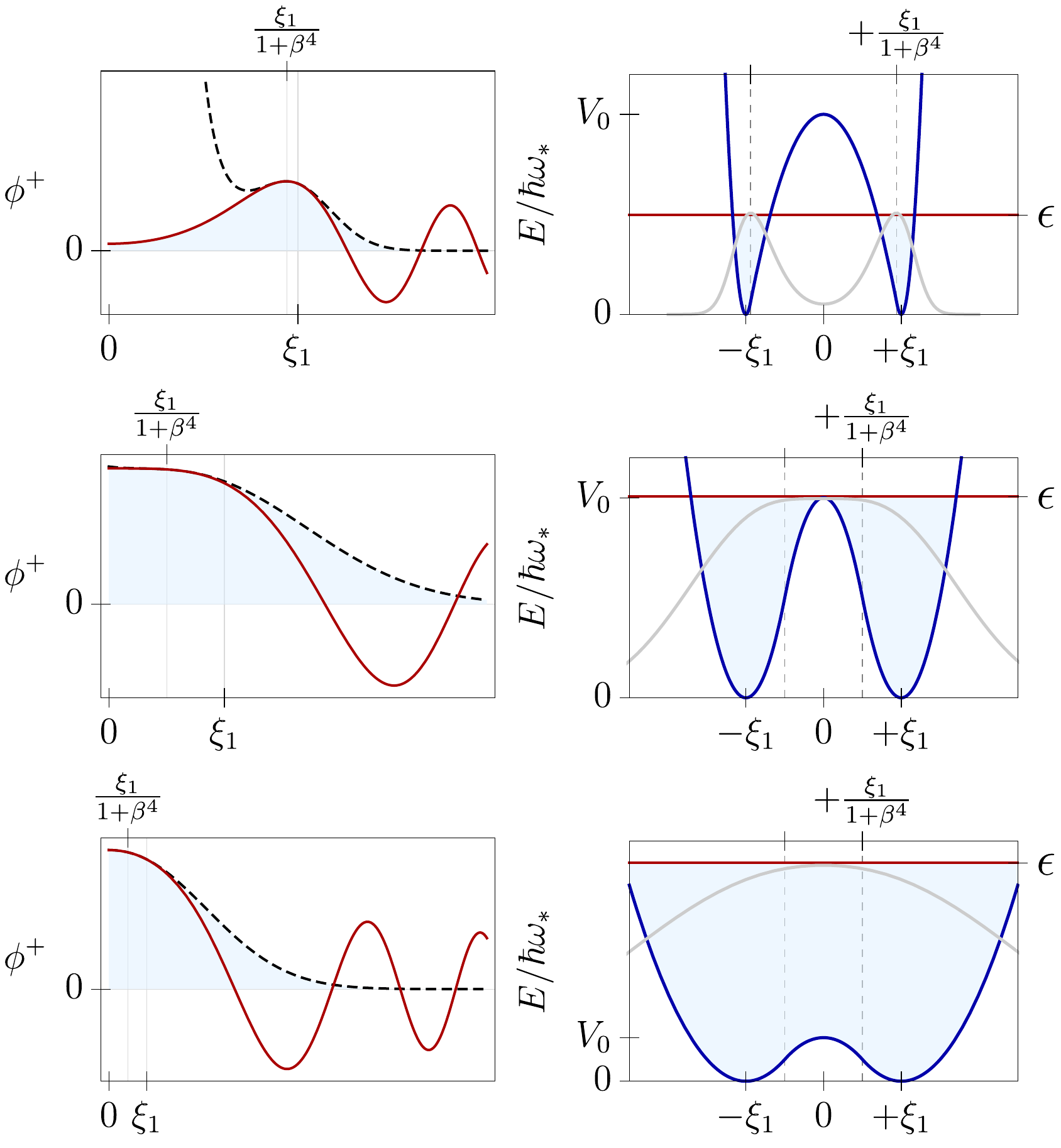}
	\caption{\label{grid}
		In the left column the ground state $\phi^{+}$ valid at the summit $\xi = 0$, a Kummer function from eqn.\eqref{invertedGS} (red curve), is joined to the parabolic cylinder function of eqn. \eqref{globalGSES} (black dashed curve) valid within the well. The area under the correct piecewise solution for $\phi^{+}$ is shaded blue. The barrier height $V_0$ is proportional to the harmonic mean of the squared frequencies, eqn.\eqref{barrier-height}. The particular parabolic cylinder function is chosen that decays to zero at $\xi \sim \infty$. Three cases are considered in turn, vacuum energies $E^{+}$ below, equal to, and above the barrier of height $V_0$,  indicated in the right column, with rows corresponding to $\{\xi_1, \beta\} \mapsto \{2.5, 0.5 \},  \{1.07 ,1 \},  \{0.5 ,1 \}$ values. Recall that $\pm \xi_1$ is the location of each well minimum, the ratio of well width to barrier width is $\beta = \sigma_{1}/\sigma_{*} = \sqrt{\omega_{*}/\omega_1}= [V''(\xi = 0)/V''(\xi= \xi_1)]^{1/4}$, and $\pm \xi_1/ (1+ \beta^4)$ are the `seams'.}
\end{figure}

These can be solved via parabolic cylinder functions, remembering the  boundary conditions that  $\phi^{-}$  and $d\phi^{+}/d \xi$ are necessarily zero at the origin $\xi = 0$ (because $\phi^{+}$ is an even function and $\phi^{-}$ is odd). 
The unnormalized ground state/excited state solutions at the barrier summit are\cite{ford1959quantum,connor1968analytical}
\begin{equation}\label{summitGSES}
\phi_0^{\pm} = D_{-1/2 +i \delta^{\pm}}\left[ (1-i) \xi \right] \pm  D_{-1/2 +i \delta^{\pm}}\left[- (1-i) \xi \right]  
\end{equation}
decomposed as an even/odd superposition of parbolic cylinder functions $D_\nu(a z) \pm D_\nu(-a z) $ respectively, with complex $a, \nu$. The ground state is expressible as a Kummer confluent hypergeometric function,  $_1F_1$ :
\begin{equation} \label{invertedGS}
\phi^{+}_0 = e^{-\frac{i \xi ^2}{2}} \, _1F_1\left(\frac{1}{4} (1-2 i
\delta^{+} );\frac{1}{2};i \xi ^2\right) ,
\end{equation}
a real function of the scaled coordinate $\xi$ and eigenvalue $\delta^{+}$ that we have pinned to  $\phi^{+}_0(0) = 1$. 

Moving on to the wavefunctions at  the well minima (and beyond to $\xi \sim \pm \infty$) -- in the scale-free setting these are parabolic cylinder functions\cite{merzbacher1970quantum}. Examining first the well centered on $\xi = \xi_1$,  the Schr\"odinger equation for the ground state takes a form associated with Weber:
\begin{equation}
-\frac{d^2 \phi^{+}_0}{d \xi^2} +\left[\frac{(\xi-\xi_1)^2}{\beta^4} - 2 \epsilon\right]\phi^{+}_0  = 0.
\end{equation}
The vacuum energy is $\epsilon = (\nu +1/2)/\beta^2$ and for an isolated well the modified eigenvalues $\nu$ are all non-negative integers. In the case of the ground state $\nu =0$, the associated eigenfunction would be a simple Gaussian profile, the lowest order Hermite function, as one would expect of a simple harmonic oscillator. For a double well with very large separation $\xi_1 \gg  \beta$ the ground and first excited states are well-approximated by an even or odd superposition of two Gaussian components centered on each well. This becomes invalid as the wells are allowed to approach one another. (The double-Gaussian ansatz is the wrong choice of orthonormal basis to span the two-dimensional subspace of $\phi^{\pm}_0$.) For finite-width barriers, tunneling causes the eigenvalues to shift away from integer values and the ground state in proximity to the wells becomes again a parabolic cylinder function $D_\nu$, this time with real non-integer $\nu$:

\begin{equation}\label{globalGSES}
\phi^{\pm}_0\left(\xi > \frac{+\xi_1}{1+\beta^4}\right)= 
D_{\beta ^2 \epsilon -\frac{1}{2}}\left[\frac{\sqrt{2}
	\left(\xi -\xi _1\right)}{\beta }\right]
\end{equation}

Of the possible solutions to the Weber equation this particular form uniquely approaches zero in the limit $\xi \sim \infty$, a necessary boundary condition for normalization, to make the wave-function square integrable. For the left well we choose the solution that approaches zero as $\xi \sim -  \infty$, actually the prior solution reflected in the $y$-axis:

\begin{equation}\label{globalGSES2}
\phi^{\pm}_0\left(\xi < \frac{-\xi_1}{1+\beta^4}\right) = 
\pm D_{\beta ^2 \epsilon -\frac{1}{2}}\left[\frac{-\sqrt{2}
	\left(\xi +\xi _1\right)}{\beta }\right]
\end{equation}
This is a guaranteed independent solution to Weber's equation, as long as $\beta ^2 \epsilon -\frac{1}{2}$ is not a non-negative integer. For the double well we have constructed, the parabolic regions were stitched to the inverted parabola at $\xi = \pm \xi_1/(1+\beta^4)$ in the scale-free coordinates. We must therefore join our two ground (or excited) state solutions also at these locations. The two states above are solutions for different regions of the potential, they do not exist in superposition, unlike eqn.\eqref{summitGSES} near the summit, where a superposition of parabolic cylinder functions was necessary to achieve the required even/odd parity about $\xi = 0$. (Note the sign change in eqn.\eqref{globalGSES2} for the excited state so it can be joined on to the odd parity solution of eqn.\eqref{summitGSES}.)

Confining attention to $\xi>0$ there are two conditions that allow our solutions to be matched, associated with the continuity of both the wavefunction and its derivative at the join. Matching eqn.\eqref{summitGSES} to the parabolic cylinder function eqn.\eqref{globalGSES2} provides the relative amplitude of the symmetric state scattered off the potential summit. Then also matching gradients at the join is only possible for a discrete set of energy eigenvalues, the lowest of which is the vacuum energy $\epsilon$.

We are most interested in the gap $\delta^{-}\leftrightarrow \delta^{+}$ so let's multiply equations \eqref{gap-pair-eqns} respectively by $\phi^{-}_0$ and $\phi^{+}_0$, and then subtracting one from the other, producing:
\begin{equation}
\frac{d^2 \phi^{-}_0}{d \xi^2} \phi^{+}_0 - \frac{d^2 \phi^{+}_0}{d \xi^2} \phi^{-}_0 + 2 \phi^{+}_0 \phi^{-}_0 (\delta^{+} -\delta^{-}) = 0 .
\end{equation}
Next we may integrate by parts, using the result:
\begin{equation}
\int_{A}^{B} \left(\psi_1 \psi''_2 - \psi_2 \psi''_1 \right) d \xi = \left.( \psi_1 \psi'_2 - \psi_2 \psi'_1) \: \right|^{B}_{A}
\end{equation}
With integration limits $\xi \in  [0,\infty]$, and recalling that $\phi^{-}(0) = 0$, and that for normalization purposes we expect $\phi^{\pm}(\infty) = 0$, we arrive at
\begin{equation}\label{gapeqn}
\frac{\Delta}{\hbar \omega_{*}} =| \delta^{-} - \delta^{+}|  = \frac{\phi^{+}_0(0) \frac{d \phi^{-}_0}{d \xi}(0) }{2 \int_{0}^{\infty} \phi^{+}_0 \phi^{-}_0 d \xi}
\end{equation}
where the denominator is the semi-overlap of the ground and excited states.

\begin{figure*}
	\includegraphics[width=6.4in]{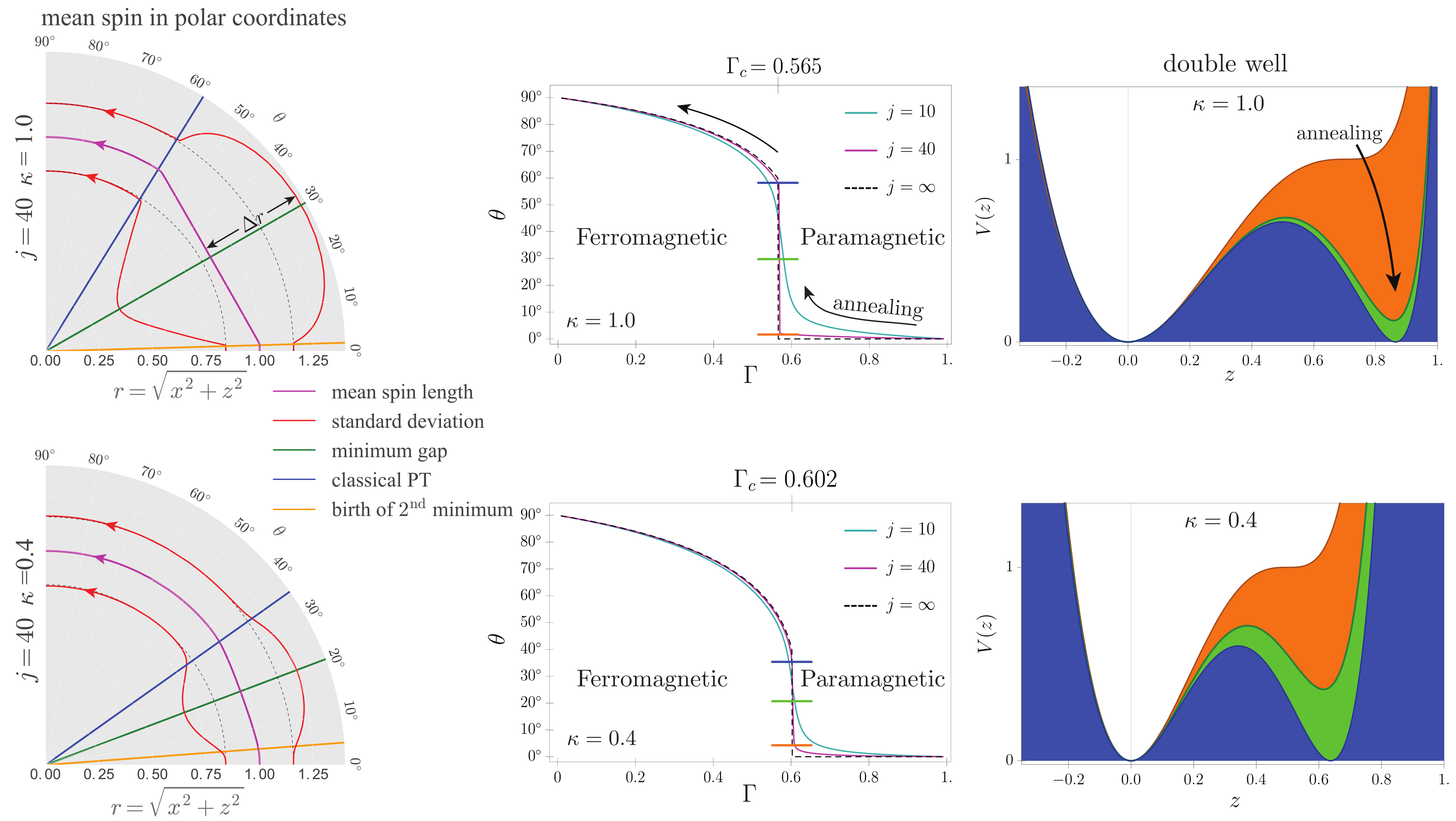}
	\caption{\label{polarplot2}
		Mean spin for the $3$-spin model has components $\langle \vec{J} \rangle = \langle\Psi |  \{ \hat{J}_{x}, \hat{J}_{y},  \hat{J}_{z} \}|  \Psi \rangle $, normalized by the principal quantum number $j$. In polar coordinates the magnitude $r = | \langle \vec{J} \rangle |/j$ and angle $\theta = \arctan \{ \langle \hat{J}_{z} \rangle /  \langle \hat{J}_{x} \rangle \}$. The uncertainty is the standard deviation (red curves) calculated from the quantum variance, $j \Delta r = \sqrt{\Delta^2 \hat{J}_{x} +\Delta^2 \hat{J}_{y}+\Delta^2 \hat{J}_{z} }$. Dashed quarter circles in the polar plots correspond to uncertainty of a classical spin (a spin coherent state), constant at $1/\sqrt{j}$, in these units. Most of the quantum behaviour is confined to a small parameter region $\Delta \Gamma$ close to the classical first order phase transition (PT) value, $\Gamma_c$. Outside the region bounded by orange/blue lines the system behaves as a large rotating spin with $r=1$, and $\Delta r = 1/\sqrt{j}$. Within this `quantum' boundary, however, the vector is quite non-classical: $r<1$ and $\Delta r > 1/\sqrt{j}$. The  orange/blue boundary lines in the polar plot correspond to the birth of the second potential minimum and the classical phase transition, respectively. At $j=40$ the appearance of the minimum spectral gap (green line) occurs at an angle almost exactly bisecting the quantum sector. The associated angle coincides with maximum uncertainty, $\Delta r$. Within the boundary the spin vector describes almost a straight line chord (magenta) perpendicular to the minimum gap  `event' angle (green line). The phase transition is  seen to be  `softer' at lower spin number, e.g. $j=10$  (cyan curves of middle column). The $j = \infty$ line (black dashed curves of middle column)  tracks the global minimum of $V(z)$ exactly   -- this is the thermodynamic limit where the vacuum (kinetic) energy vanishes. The right-most plots illustrates the shape of the double well for $\kappa \mapsto \{1.0,0.4\}$ at the birth of the second minimum (orange) and at the classical phase transition (blue) and for the minimum spectral gap in the case $j=40$ (green). Observe that at the minimum gap the potential is \emph{not} symmetric, nor the wells of equal depth.}
\end{figure*} 

\section{Analysis of the Phase Transition in the Catalysed 3-Spin Model}

Often, phase transitions in quantum spin systems are modeled using a mean field model, where the ground state is represented as a product state of $n$ spins (a spin coherent state) that tracks the potential minimum during the annealing process. The validity of such a description  can be investigated, both in the neighborhood of the phase transition, and far from it. Without the possibility of entanglement the model of a large rotating classical pointer falls short, in particular in the transition region, as evidenced by FIG.\ref{polarplot2}. Classically, one usually defines the phase transition as occurring when the two well minima are of equal depth, but in the quantum case we should instead nominate the minimum gap location on the annealing landscape, which itself depends on the number of spins $n$. Additional complications in the $3$-spin model are: mass that varies as a function of the well location, and the fact the double well is asymmetric at the minimum gap.

\begin{figure}
	\includegraphics[width=3.0in]{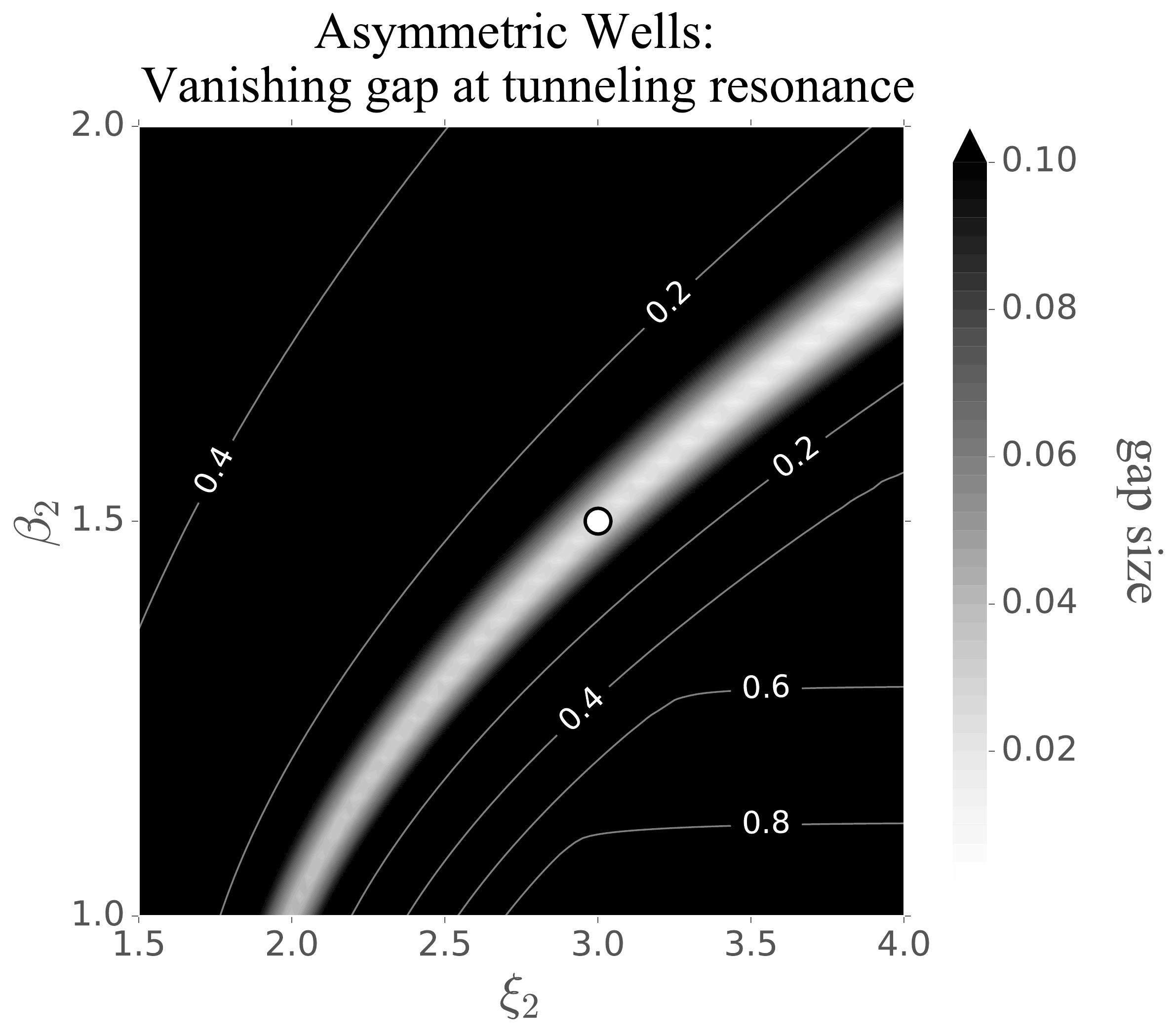}
	\caption{\label{no_rayleigh}
		Asymmetric double-well: Here is illustrated the case that the left well is outside the Rayleigh boundary: $\{ \xi_1, \beta_1\} \mapsto \{ 3.0,1.5\}$ (white circle marker). Then one may employ WKB-like methods; the gap can become so small on resonance to be dictated by the potential shape and structure  away from the parabolic extrema\cite{vybornyi2014tunnel}. In such a case the piecewise-parabolic model loses its generality; errors or simplifications in the description of the potential have greater magnitude than the spectral gap calculated via this potential. The light-shaded channel identifies the resonance condition for $\{ \xi_1, \beta_1, \xi_2, \beta_2\} $; an exponentially small minimum gap where one can expect  a first-order phase transition via slow tunnelling.}
\end{figure}

\section{Catalysed 3-Spin model in the $\kappa \ll 1$ Limit} \label{appendix_kappa}

In the regime $0< \kappa \ll 1$ near the phase transition, the annealing ratio at the quantum boundary (birth of the second minimum) is 
\begin{widetext}
\begin{align} \label{gamma_zero}
\gamma_0 = \frac{\sqrt{169 \kappa ^4-172 \kappa ^3+78 \kappa ^2+8 \kappa -2 (\kappa -1) (\kappa 
		(19 \kappa -2)+1)^{3/2}-2}}{6 \kappa }  \: .
\end{align}
\end{widetext}
The asymptotic expansion to second order of this expression is:
\begin{equation}
\gamma_0 = 2 (1- \kappa) + \frac{9}{4} \kappa^2 + O(\kappa^3).
\end{equation}
Interestingly, this is not a convergent series for all $\kappa \in [0,1]$, cubic and higher terms are ignored at our peril. The hidden second order phase transition (occurring when the paramagnetic minimum at the origin $z=0$ becomes a maximum) occurs precisely at $\gamma_{2} = 2 (1- \kappa) $, which is the same as $\gamma_0$ to to first order in $\kappa$. The quantum phase transition must occur between $\gamma_0$ and $\gamma_2$, and therefore must be $O(\kappa^2)$ or smaller.

We choose to parameterize this sweet spot as 
\begin{equation}
 \gamma_0 - \gamma_{*} = \delta \gamma  =  \left[ \frac{x \kappa}{2} \right]^2
\end{equation}
where $x \lesssim 1.07$ is a small parameter. (Value $x=1.07$ corresponds to the classical phase transition, at which the wells have equal depth.)

The well separation within this quantum regime is
\begin{equation}
z_1= \frac{(3+x)\kappa}{2}+ O(\kappa^{2}) \; ,
\end{equation}
and the distance to the maximum from the minimum at the origin is
\begin{equation}\label{zstar}
z_{*}=  \frac{(3- x)\kappa}{2}+ O(\kappa^{2}) \; ,
\end{equation}
The barrier summit and ferromagnetic ground state are therefore separated by
\begin{equation}\label{zminzstar}
z_1 -z_{*} = \kappa x + O(\kappa^{2}) \; ,
\end{equation}
zero at $\gamma = \gamma_0$ or $x=0$, where these two extrema merge.

The barrier, or potential difference to the summit from the minimum at $z=0$, is polynomially dependent on $\kappa$:
\begin{align}\label{V0}
V_0 =  \frac{\kappa^4}{128}(3-x)^3(1+x) +O(\kappa^5)
\end{align}
On the other side of the summit, the second minimum is below the maximum by an energy:
\begin{equation}
V_{00} =  \frac{x^3 \kappa^4}{8} +O(\kappa^5) \: .
\end{equation}

In the same limit, the characteristic frequencies associated with the quadratic extrema of $V(z)$ are:
\begin{subequations}
	\begin{align}
	\omega_{0} &\approx \frac{9-x^2}{8} \kappa^2 +O(\kappa^3) \;  , \\
	\omega_{1}& \approx  \frac{\sqrt{3x} }{4} (3 + x )  \: \kappa ^{2}   +O(\kappa^3) \;  ,  \\
 \omega_{*}& \approx \frac{\sqrt{3x} }{4} (3 - x )  \: \kappa ^{2}   +O(\kappa^3)  \; .  \label{freqs}
	\end{align}
\end{subequations}
To summarize, in the small $\kappa$ limit that coordinate distances,  characteristic frequencies and potential barrier scale  with the non-stoquastic driver as $\mapsto \{\kappa, \kappa^2, \kappa^4 \}$.  FIG.\ref{barrierk} shows at the classical phase transition point how the barrier size and shape change with $\kappa$. (The implicit $\Gamma$ values are chosen to maintain the wells at equal depth as $\kappa$ is varied.) 
At this point we underscore some of the subtleties of this $3$-spin model, namely the asymmetry of the double well compounded by a position-dependent mass variable, eqn.\eqref{inverse_mass}. The WKB method has been discussed in the literature for tunneling under a barrier between asymmetric wells\cite{halataei2017tunnel,rastelli2012semiclassical} though we find no previous work that discusses the coherent catalysis limit of  a `low barrier', $\hbar \omega/2 \sim V_0$ (where lowest order WKB truncation will likely fail). We simplify the position-dependent mass challenge by taking delta samples of the mass at the potential minima. This is a reasonable assumption for narrow wells and  $j \gg 1$, but less accurate as wells become more shallow and merge. Even so, the characteristic frequency at a quadratic minimum will be $\omega(z) = \sqrt{V''(z) / m(z)}$ and the quantum ground state is `heavier' within the ferromagnetic phase centered on $z_1>0$ than at the $z=0$ (paramagnetic phase) ground state. Recall that $\beta$ is a ratio of frequencies, and not a function of $\hbar$: 
\begin{equation} \label{beta}
\beta = \frac{\sigma_1}{\sigma_{*}} \approx \sqrt{\frac{m_{*} \omega_{*}}{m_1 \omega_1}}
\end{equation}
On the other hand, $\xi_1 =\Delta z/ \sigma_{*} = \Delta z \sqrt{m_{*} \omega_{*}/ \hbar}$. This is where the energy scale will enter --  in terms of parameter $\hbar = 1/j$.

We shall also need the $\kappa \sim 0$ asymptotic expression for the variable masses:
\begin{subequations}\label{var_mass_eqn}
	\begin{align}
	\frac{1}{m_{0} } &= \omega_0 \approx  \frac{9-x^2}{8} \kappa^2 +O(\kappa^3) \; ,  \\
	\frac{1}{m_{1} }& \approx \frac{3}{4} (3 +x ) \kappa ^2  +O(\kappa^3) \;   , \\
	\frac{1}{m_{*} }& \approx\frac{3}{4} (3 -x )  \kappa ^2 +O(\kappa^3) \;   .
	\end{align}
\end{subequations}
Notice that $m_0 \omega_0 = 1$, and therefore $\sigma_0 = 1/ \sqrt{j}$. For the Rayleigh limit, powers of $\kappa$ balance on both sides of $z_{*} \approx\sqrt{\hbar/(m_{0} \omega_{0})}$ only if $\hbar= O(\kappa^2)$, since $z_{*} = O(\kappa)$,  eqn.\eqref{zstar}. Re-ordering the terms, demanding $\xi_1 \sim \beta$ leads to a scaling law at the maximum minimum spectral gap (saddle):
\begin{equation}
\kappa_c  \sim O\left(1/\sqrt{j} \right) \; .
\end{equation}
This result is employed to derive the asymptotic scaling of the saddle spectral gap for the 3-spin in section \ref{3spin}.

Other useful asymptotic expressions for $\kappa \ll 1$ in the quantum region between the birth of the second minimum and classical phase transition point are:
\begin{equation} \label{beta2}
\frac{1}{\sigma_{1}}  \approx\frac{1}{\sigma_{*}}=\sqrt{j} \left( \frac{x}{3}\right)^{1/4} \left(1 - \frac{9-x^2}{8 x^2} \kappa \right) + O(\kappa^{2})
\end{equation}
The ratio deviates from unity only slightly:
\begin{equation}
\beta_2 = \frac{\sigma_1}{\sigma_{*}}  =1-\frac{15}{8} x \kappa ^{2} + O(\kappa^{3})  \label{ratio_beta}
\end{equation}
The scaled distance from the barrier summit to the ferromagnetic minimum is
\begin{equation}
\xi_2 = \frac{z_1 - z_{*}}{\sigma_{*}} = \sqrt{j} \left[ \frac{x^{5/4}}{3^{1/4}}  \kappa + O(\kappa^{2}) \right] 
\end{equation}

The remaining two parameters are:
\begin{align}
\xi_1 &= \frac{z_{*}}{\sigma_{*}} =\sqrt{j} \left[ \left( \frac{x}{3}\right)^{1/4} \left(\frac{3-x}{2} \right) \kappa + O(\kappa^{2}) \right]\\
\beta_1  &=  \left( \frac{x}{3}\right)^{1/4} + O(\kappa)
\end{align}

If we apply the ground-state resonance condition of section \ref{resonance_section} to our scale-free model, this produces $x =1$. FIG.\ref{antirayleigh} presents the asymmetric well with $x=1$. The parameters above then map onto eqn. \eqref{rayleigh_answer} in the main text.

\section{Quantum 2-Spin: Lipkin Meshkov Glick } \label{LMG}

The quantum $3$-spin will prove challenging to implement experimentally. In contrast, the $2$-spin, the simplest $p$-spin model, is an isotropic variant of the Lipkin Meshkov Glick (LMG) model introduced by Fallieros in 1959 in Ref.\onlinecite{Fallieros1959} to describe the nuclear physics of Oxygen, and re-visited  by Lipkin and collaborators in Ref. \onlinecite{lipkin1965validity}. It is an Ising model with infinite range interactions; however, with $2$-local rather than $3$-local interactions the implementation on an experimental quantum annealer (such as that of Ref.\onlinecite{johnson2011quantum}), may be relatively manageable. Current devices have up to $n = 4000$ qubits, but do not implement the fully connected graph required of the LMG model, (physical spins are topologically constrained to couple to nearby spins). Their partially connected architecture does, in fact, admit simulations of fully-connected models of smaller ensembles, via a process called `embedding'. 

\begin{figure}
	\includegraphics[width=3.in]{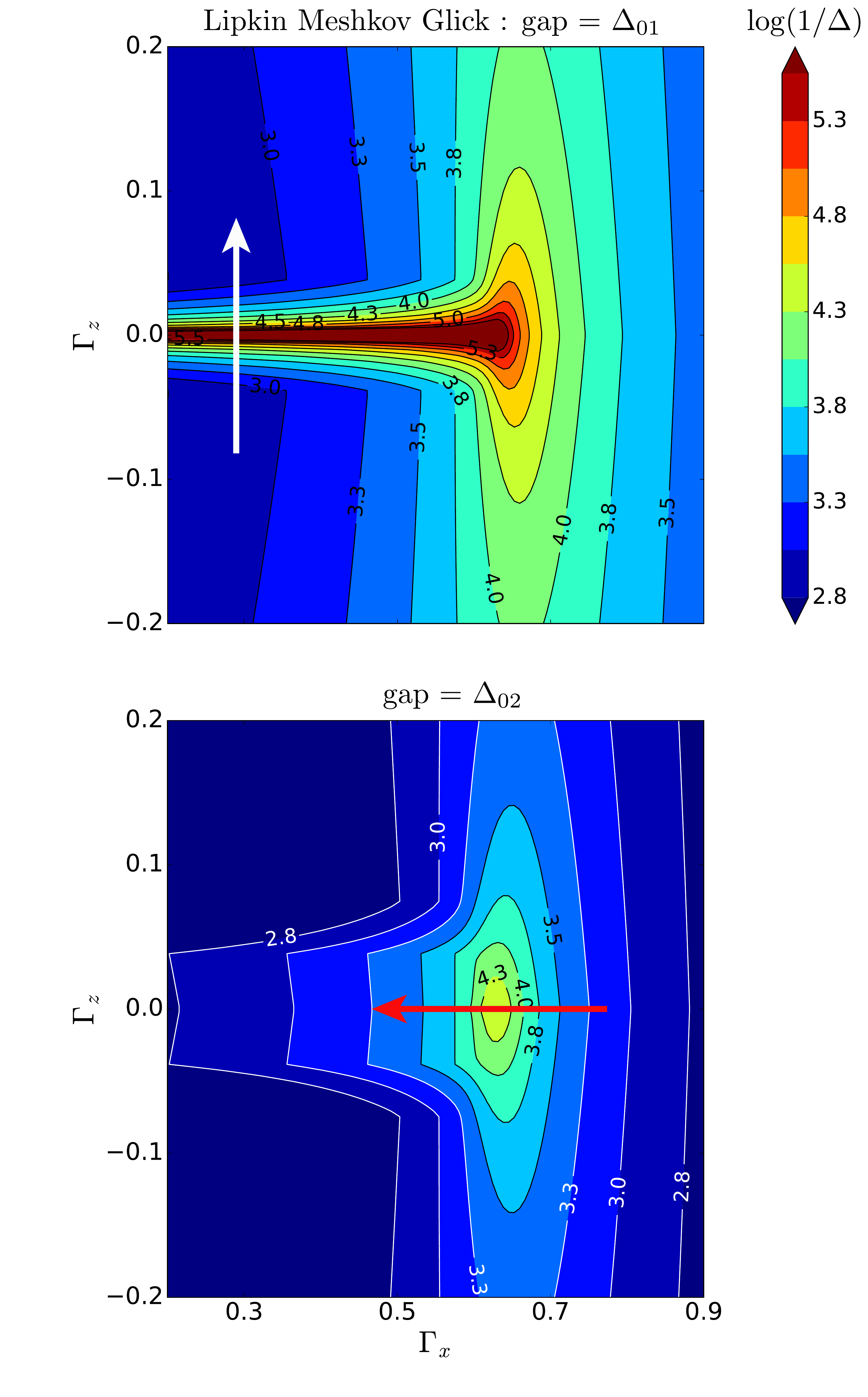}
	\caption{\label{LMG_gap}
		The  Lipkin-Meshkov-Glick (LMG) model with transverse and longitudinal field parameters $\Gamma_x, \Gamma_z$ respectively, exhibits a relatively gentle second order phase transition we explored in Ref.\onlinecite{durkin2016asymptotically} (lower plot) . It occurs for an annealing schedule that follows  the line of $\Gamma_z = 0$ from $\Gamma_x > 2/3$ to $\Gamma_x <2/3$  (direction of red arrow). The minimum gap $\Delta_{02}$ near $\Gamma_z = 0, \Gamma_x = 2/3$ is polynomial in $n$ -- transitions between ground and first excited state $\Delta_{01}$ are forbidden by parity along $\Gamma_z = 0$.  In contrast, an annealing schedule for which $\Gamma_z \neq 0$ will not respect parity and the relevant gap is $\Delta_{01}$ (upper plot). A schedule crossing this zero longitudinal field line (direction of white arrow) must undergo a first order phase transition if $\Gamma_x \lesssim 2/3$. Gap $\Delta_{01}$ is then exponentially small when crossing $\Gamma_z = 0$. The landscape above is for a $j=25$ spin ensemble.}
\end{figure}

For LMG we dispense with the anti-ferromagnetic driver $+(1-\kappa) \hat{J}_x^2$.  Instead there may exist some longitudinal field component in addition to the transverse one, represented by control parameters $\Gamma_{z,x}$ respectively:

\begin{equation} \label{spinH}
\hat{H}_{\text{LMG}} = - \Gamma_x \frac{\hat{J}_x}{j} -  (1- \Gamma_x) \left[ (1 - |\Gamma_z|) \frac{\hat{J}_z^2}{j^2} +   \Gamma_z  \frac{\hat{J}_z}{j}  \right] \; \;  
\end{equation}
We  previously studied the behaviour of this model during quantum annealing for the $\Gamma_z = 0$ setting\cite{durkin2016asymptotically}, it undergoes a second order phase transition close to $\Gamma_x = 2/3$ where the unimodal ground state smoothly and continuously bifurcates into a bimodal Schr\'odinger cat state, eventually becoming a GHZ state\cite{greenberger1989going}. The associated minimum gap at the phase transition is polynomially small, the symmetry of this model means that adiabatic transitions are forbidden from the ground to first excited state due to their opposite parity; the relevant minimum gap along the contour $\Gamma_z = 0$ in parameter space is actually $\Delta_{02} = E_2- E_0 \sim n^{-4/3}$. For non-adiabatic (e.g. thermal) transitions $E_0 \mapsto  E_1$, or the first order annealing transitions where $\Gamma_z$ is non-zero and switches sign, that gap also scales $\propto n^{-4/3}$ near $\Gamma_x \approx 2/3$ before becoming exponentially small in the tunnel-splitting limit characterized by the Gamow factor when $\Gamma_x \ll  2/3$. All of this is illustrated in the contour plots of FIG.\ref{LMG_gap}. 

The LMG continuous potential function becomes
\begin{equation}
V^{[\text{LMG}]}(z)  = - \sqrt{1-z^2}   - \left( \frac{(1-|\Gamma_z|) z^2}{\gamma_x} +  z \: \Gamma_z  \right) + O(\hbar) \; .
\end{equation}

For zero longitudunal field $\Gamma_z$ and within the ferromagnetic phase $\Gamma_x < \Gamma_0$ the potential above is a beautifully symmetric double well, with a barrier height fully controlled by $\Gamma_x$. In term of the ratio $\gamma_x = \Gamma_x/(1-\Gamma_x)$ the barrier is:
\begin{equation}\label{LMG_V0}
V_0^{[\text{LMG}]} = \frac{1}{\gamma_x} + \frac{\gamma_x}{4} -1 .
\end{equation}
This very regular potential with a simple analytical form provides a perfect setting in which to examine the mechanism of vacuum delocalization. For $\gamma_x$ increasing through the critical point $\gamma_x \mapsto \gamma_c = 2$, the potential barrier is completely razed to a flat-bottomed quartic profile $\sim z^4$ that evolves further into a single quadratic minimum centered at the origin $z=0$ for $\gamma_x >2$. The classical transition at $\gamma_x = 2$ is described by a green contour line in the right-side plots of FIG.\ref{PT}. In the small $\gamma_x$ ferromagnetic phase  (double well potential) one may introduce asymmetry or bias in the potential by a  (positive or negative) longitudinal field of $|\Gamma_z| \ll 1$. This  will lower one well minimum with respect to the other and the adiabatic ground state will lose its fragile superposition state. It shifts completely to being a spin coherent state pointed at the deeper well. 
\begin{figure}
	\includegraphics[width=3.in]{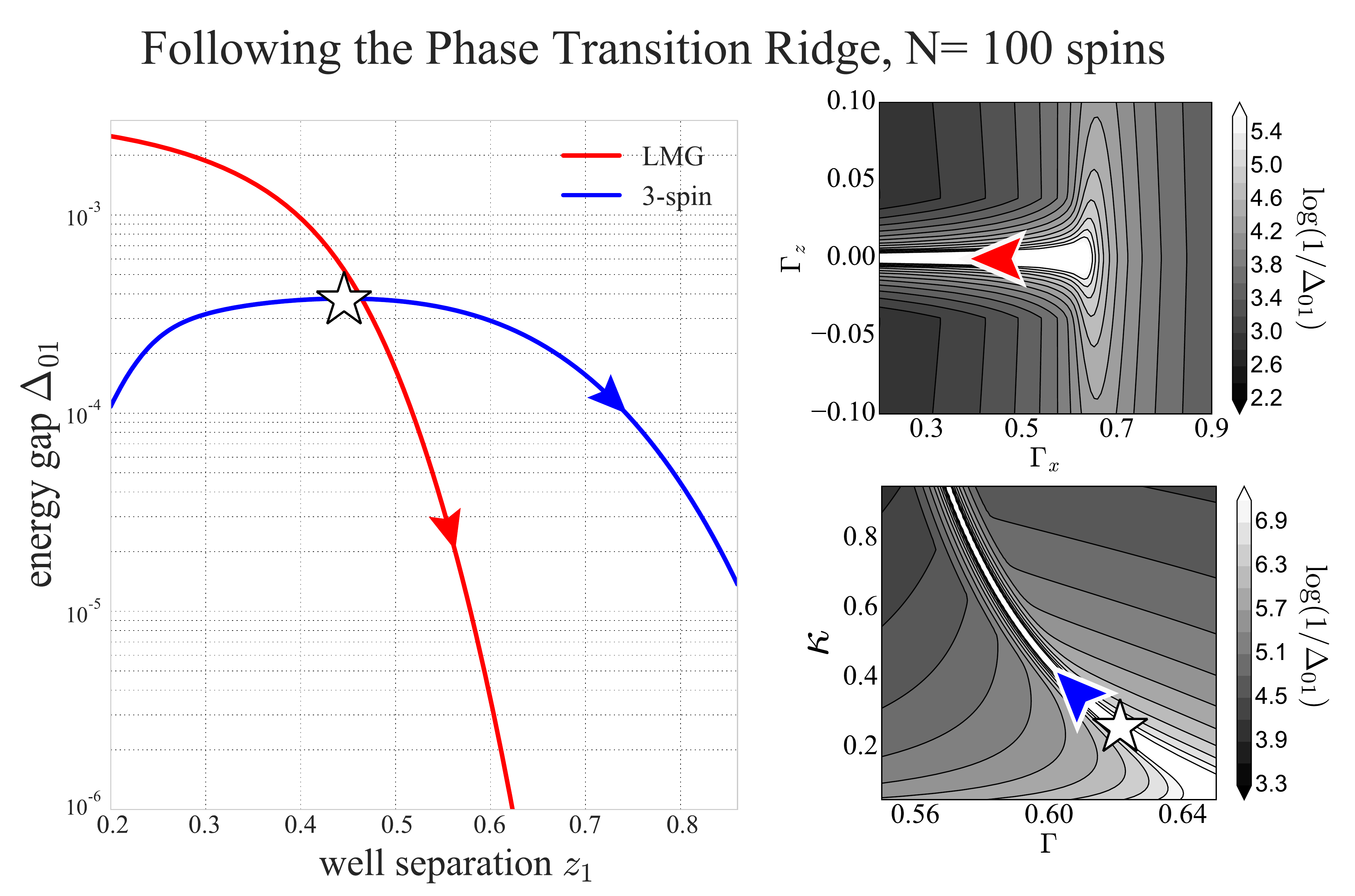}
	\caption{\label{ridge}
		For both Lipkin-Meshkov-Glick (LMG, red curve) and quantum $3$-spin model (blue curve), increasing the separation of the well minima $z_1$ leads to exponentially small minimum gaps, which occur along the `ridge' of the phase transition (contour plots to the right show these ridges in parameter space and the direction along which they are traversed as $z_1$ is increased). The $3$-spin model is distinctive in that the minimum gap (blue) goes through a maximum in  $z_1$, indicating a saddle point of `optimal catalaysis' (white star marker). Exploring the ridge in the reversed direction now, from top left to bottom right (lower right panel), the well separation continues to decrease with decreasing $\kappa$, but the inverse mass $1/m \propto \kappa^2$. There is apparently a competition between the quantum `particle' becoming more confined at lower $\kappa$, but heavier at the same time. The increasing mass eventually wins at very low $\kappa$, causing the spectral gap to shrink again.}
\end{figure}

\begin{figure}
	\includegraphics[width=2.4in]{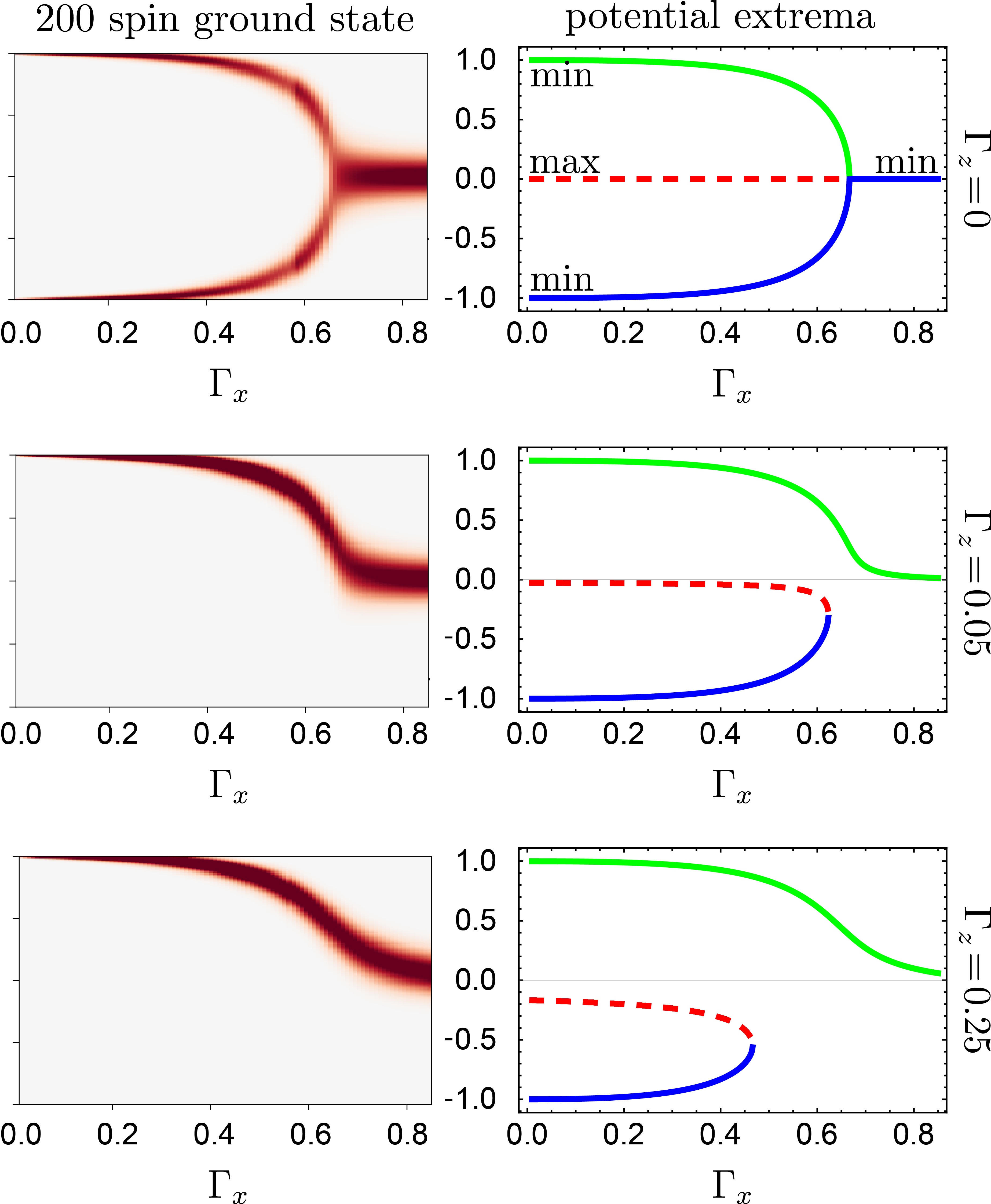}
	\caption{\label{LMGxz}
		For longitudinal field  $\Gamma_z \mapsto \{ 0,+0.05,+0.25\}$ the ground state amplitude density (found numerically) and potential extrema (derived analytically) are shown in the left and right columns respectively. For the right column, the locus of the potential maxima in coordinate space $z$ is indicated (dashed red line), with the exterior minima are shown in blue (shallow) and green (deep). As the barrier maximum is lowered with increasing $\Gamma_x$, for a symmetry breaking $\Gamma_z >0$ one of the minima collides with the now off-center maximum. Only the green minimum remains, moving towards the origin $z=0$ as $\Gamma_x$ is increased further. Changing $\Gamma_z$ to $-\Gamma_z$ just reflects the diagrams through the origin (top to bottom). In the lower figure then the minimum at $z \approx -1$ would become the true minimum, and the (now false) ground state at $z= +1$ will tunnel through the intervening barrier (red dashed maximum line), unless the barrier is small enough, Eqn.\eqref{LMG_V0}, for vacuum delocalization to occur.}
\end{figure}

Then by reversing the bias of $\Gamma_z$ the ground state will have to tunnel across the large intervening barrier from the false minimum to the true minimum. A magnetization measurement should be able to record the characteristic time for the population inversion to occur. The extrema of the potential energy surface are plotted in FIG.\ref{LMGxz}.

When  $\gamma_x$ is increased to lower the intervening barrier to a fixed height, (the analogue of non-stoquastic catalysis in the p-spin model), one may observe directly a crossover in the characteristic magnetization `switching time', (population transfer from the left to the right well) as $\Gamma_z$ is varied from slightly negative to slightly positive. Increasing the transverse field, the $\gamma_x \lesssim 2$ or $\Gamma_x \lesssim 2/3$  regime  (wells start to coalesce as the potential barrier between them shrinks) should permit a polynomial gap for a range of $\Gamma_x$ allowing the ground state energy to delocalize close to the barrier summit, before the wells completely coalesce at $\Gamma_x = 2/3$. The change in the rate of population inversion from exponentially slow to rapid polynomial time scales  should be apparent and measurable in the $\Gamma_x \lesssim 2/3$ regime. Because parameter $1/j$ plays the role of an effective $\hbar$ when transforming to a particle in a potential, smaller ensembles exhibit more `extravagantly' quantum effects, e.g. magnified vacuum energies, allowing increased mobility across barriers without (exponentially slow) tunneling.

\end{document}